\newif\ifnowc
\nowctrue
\documentclass[aps,reprint]{revtex4-1}
\usepackage{scrextend}
\usepackage{bibentry}
\usepackage{amstext}
\usepackage{bm}            %serve per le subequazioni
\usepackage{amsmath} 
\usepackage{amssymb}            %serve per il simbolo "marchio registrato", \circledR
\usepackage{graphicx}           %serve per le figure eps, ps etcyy
\usepackage{latexsym}
\usepackage{color}
\usepackage{stackengine}
\usepackage{subfigure}
\usepackage{natbib}

\newcommand{\bibliopath}{/Users/carollo/Docs/Articles&Manuscripts/}
\newcommand{\Tr}{\text{Tr}}
\newcommand{\tr}{\text{tr}}
\newcommand{\Det}{\text{Det}}
\newcommand{\Cov}{\text{Cov}}
\newcommand{\Ad}{\text{Ad}}
\newcommand{\Pf}{\text{Pf}}
\newcommand{\ket}[1]{|#1\rangle}
\newcommand{\bra}[1]{\langle #1|}
\newcommand{\bk}[1]{\langle #1 \rangle}
\newcommand{\HH}{\mathcal{H}}
\newcommand{\one}{\mbox{$1 \hspace{-1.0mm}  {\bf l}$}}
\newcommand{\vv}[1]{{#1}}
\newcommand{\V}[1]{\text{vec}({#1})}
\newcommand{\QED}{\hfill\ensuremath{\blacksquare}}
\newcommand{\Title}{Uhlmann curvature in dissipative phase transitions}

\renewcommand{\Re}{\mbox{\bf Re}}
\renewcommand{\Im}{\mbox{\bf Im}}

\begin{document}
\title{\Title}
\author{ Angelo Carollo$^{\diamond \ddag}$, ,
Bernardo Spagnolo$^{\diamond \ddag \dag}$ and Davide Valenti$^{\diamond}$}
\affiliation{$^{\diamond}$Department of Physics and Chemistry, University of Palermo, Viale delle Scienze, Ed. 18, I-90128 Palermo, Italy\\
$^\ddag$Radiophysics Department, Lobachevsky State University of Nizhni Novgorod, 23 Gagarin Avenue, Nizhni Novgorod 603950, Russia\\
$^\dag$Istituto Nazionale di Fisica Nucleare, Sezione di Catania,  Via S. Sofia 64, I-90123 Catania, Italy}
\begin{abstract}
We study the mean Uhlmann curvature in fermionic systems undergoing a dissipative driven phase transition. We consider a paradigmatic class of lattice fermion systems in non-equilibrium steady-state of an open system with local reservoirs, which are characterised by a Gaussian fermionic steady state. In the thermodynamical limit, in systems with translational invariance we show that a singular behaviour of the Uhlmann curvature represents a sufficient criterion for criticalities, in the sense of diverging correlation length, and it is not otherwise sensitive to the closure of the Liouvillian dissipative gap. In finite size systems, we show that the scaling behaviour of the mean Uhlmann curvature maps faithfully the phase diagram, and a relation to the dissipative gap is put forward. We argue that the mean Uhlmann phase can shade light upon the nature of non equilibrium steady state criticality in particular with regard to the role played by quantum vs classical fluctuations.
\end{abstract}

\ifnowc \maketitle \fi
A challenging new paradigm has recently been put forward by the discovery of novel types of quantum phase transitions (QPTs)~\cite{Sachdev2011} occurring in non-equilibrium steady states (NESSs)~\cite{Prosen2008,*Diehl2008,*DallaTorre2010,*Diehl2010a,*Heyl2013,Ajisaka2014,*Vajna2015,*Dagvadorj2015,*Bartolo2016,*Jin2016,Roy2017,*Fink2017,*Fitzpatrick2017}. A comprehensive picture and characterisation of dissipative NESS-QPT is lacking, partly due to their nature lying in a blurred domain, where features typical of zero temperature QPT coexists with unexpected properties, some of which reminiscent of thermal phase transitions.\\\indent 
A natural approach to the investigation of such a novel scenario would be to adapt tools used in the equilibrium settings. In this letter, we propose the use of the geometric phase (GP)~\cite{Bohm2003,*Berry1984}, and in particular its mixed state generalisation, the Uhlmann GP~\cite{Uhlmann1986}, to investigate NESS-QPT.
GPs, and related geometrical tools, such as the Bures metrics~\cite{Bures1969,*Uhlmann1976,*Braunstein1994}, have been successfully applied in the analysis of many equilibrium phase transitions~\cite{Zanardi2006,Zanardi2007,*CamposVenuti2007,*Gu2010,*Dey2012}. 
%The Bures metrics have been employed in thermal phase transition~\cite{Janyszek1999,*Ruppeiner1995,*Quan2009}, and QPT, both in symmetry-breaking~\cite{Zanardi2006,*Zanardi2007,*CamposVenuti2007,*Zanardi2007a,*Gu2010,Dey2012,*Kolodrubetz2013} as well as in topological phase transitions~\cite{Yang2008}. 
GPs are at the core of the characterisation of topological phase transitions~\cite{Thouless1983,*Bernevig2013,*Chiu2016}, and have been employed in the description and detection of QPT, both theoretically~\cite{Carollo2005,*Pachos2006d,*Plastina2006,*Hamma2006,*Zhu2006,*Reuter2007} and experimentally~\cite{Peng2010}. The use of GP in QPT can be heuristically understood as follows: QPT are determined by dramatic structural changes of the system state, resulting from small variations of control parameters. When approaching a criticality, two infinitesimally close states on the parameter manifold, become increasingly statistically distinguishable, i.e. their geometric-statistical distance grows. Abrupt changes in the distance are accompanied by singularities of the state space curvature, which in turn determine GP instabilities on states traversing loops in the neighbourhood of the criticality~\cite{Carollo2005,*Pachos2006d,*Plastina2006,*Hamma2006,*Zhu2006,*Reuter2007}.\\\indent
Due to their mixed state nature, the NESSs require the use of a definition of GP in the density operators domain. Among all possible approaches~\cite{Uhlmann1986,Sjoqvist2000a,*Tong2004,*Chaturvedi2004,*Marzlin2004,*Carollo2005c,*Buric2009,*Sinitsyn2009}, the Uhlmann GP~\cite{Uhlmann1986} stands out for its deep-rooted relation to information geometry and metrology~\cite{Matsumoto1997,*Hayashi2017}, whose tools have been profitably employed in the investigation of QPT and NESS-QPT~\cite{Zanardi2007,Kolodrubetz2013,*Banchi2014,*Marzolino2017}. Uhlmann holonomy and GP have been applied to the characterisation of both topological and symmetry breaking equilibrium QPT~\cite{Paunkovic2008,*Huang2014,*Viyuela2014,*Andersson2016,*Viyuela2015,*Budich2015a,*Kempkes2016,*Mera2017}. Many proposals to measure the Uhlmann GP have been put forward~\cite{Tidstrom2002,*Aberg2007,*Viyuela2016}, and demonstrated experimentally~\cite{Zhu2011}.\\\indent 
Motivated by this, we introduce the mean Uhlmann curvature (MUC) and investigate its role in the characterisation of dissipative NESS-QPT. 
The MUC, defined as the Uhlmann GP per unit area of a density matrix evolving along an infinitesimal loop,
has also a fundamental interpretation in multiparameter quantum metrology: it marks the incompatibility between independent parameters arising from the quantum nature of the underlying physical system~\cite{Ragy2016}. In this sense, the MUC is a measure of ``quantumness'' in the \emph{multi-parameter} estimation problem, and its singular behaviour responds only to quantum fluctuations occurring across a phase transition.\\\indent
We apply these ideas to the physically relevant setting of fermionic quadratic dissipative Lioviullian models, some of which show rich NESS features.~\cite{Prosen2008,Diehl2008,Eisert2010,Banchi2014,Marzolino2014,*Marzolino2017}.\\\indent 
%,which have been investigated characterised in terms of long-range magnetic order, scaling of the dissipative gap and single parameter metrology as well as information-geometrical tools~\cite{Prosen2008,Diehl2008,Eisert2010,Banchi2014,Marzolino2014,*Marzolino2017}.
\emph{The mean Uhlmann curvature}.--
The Uhlmann GP relies on the idea of amplitude of a mixed state. Given a density operator $\rho$ acting on a Hilbert space $\HH$ of dimension $n$, an amplitude is an operator $w$ satisfying $\rho=w w^{\dag}$. This definition leaves a $U(n)$ gauge freedom in the choice of $w$, as $w'=w U$, for any $U \in U(n)$, generates the same $\rho$.\\\indent
Let $\rho_{\lambda(t)}$ be a family of density matrices, with $\gamma:=\{\lambda(t)\in\mathcal{M},\, t\in[0,T]\}$ a smooth closed path in a parameter manifold $\mathcal{M}$, and $w_{\lambda(t)}$ is a corresponding path of amplitudes. To lift the $U(n)$ gauge freedom, Uhlmann introduced a parallel transport condition on $w_{\lambda(t)}$~\cite{Uhlmann1986}. For a closed trajectory $\rho_{\lambda(t)}$, initial and final amplitudes are related by a unitary transformation $w_{\lambda(T)}=w_{\lambda(0)}V_{\gamma}$. If the Uhlmann parallel transport condition is fullfilled, $V_{\gamma}$ is a \emph{holonomy}, i.e. a non-Abelian generalisation of the Berry phase~\cite{Uhlmann1986}, and reads $V_{\gamma}=\mathcal{P}e^{i\oint_{\gamma} A}$, with $\mathcal{P}$ being the path ordering operator and $A=\sum_{\mu}A_{\mu} d\lambda_{\mu}$  the Uhlmann connection one-form. The Uhlmann GP is defined as 
$\varphi^{U}[\gamma]:=\arg{(w_{\lambda(0)},w_{\lambda(T)})}=\arg{\Tr{(w_{\lambda(0)}^{\dagger}w_{\lambda(0)}V_{\gamma})}}$.\\\indent
The Uhlmann connection $A$ can be derived from the ansatz~\cite{Uhlmann1989,*Dittmann1999} 
$\partial_{\mu} w=  \frac{1}{2} L_{\mu} w - i w A_{\mu}$, where $\partial_{\mu}:=\partial/\partial \lambda_{\mu}$, and
 $L_{\mu}$'s are Hermitian operators known as symmetric logarithmic derivatives (SLDs), implicitly defined as the operator solutions of $\partial_{\mu}\rho=:\frac{1}{2}(L_{\mu}\rho+\rho L_{\mu})$~\footnote{ Unless otherwise stated, we will assume that $\rho$ is a full-rank. If $\rho$ is singular, $L_{\mu}$ and $A_{\mu}$ are not unique. However, we will show that any quantity of interest to us can be extended by continuity to singular $\rho$'s\cite{Safranek2017}}. 
%It follows also that $A_{\mu}$ are Hermitian operators obeying the transformation rule of a non-abelian gauge potential, $A\to U^{\dagger} A U+i U^{\dagger}dU$ under $w\to w U$, while $L_{\mu}$ are gauge invariant. 
The Uhlmann curvature, defined as $F_{\mu\nu}=\partial_\mu A_\nu -\partial_\nu A_\mu-i\left[A_{\mu},A_{\nu}\right]$, is equal to the Uhlmann holonomy per unit area associated to an infinitesimal loop in $\mathcal{M}$, i.e. $F_{\mu\nu}=\lim_{\epsilon \to 0} i \frac{1-V_{\gamma_{\mu,\nu}}}{\epsilon^{2}}$, where $\gamma_{\mu \nu}$ is the infinitesimal parallelogram spanned by two independent directions $\hat{e}_{\mu}\epsilon$ and $\hat{e}_{\nu}\epsilon$ in $\mathcal{M}$. We focus on the Uhlmann GP per unit area for an infinitesimal loop, i.e.
\ifnowc\begin{equation*}
 \mathcal{U}_{\mu\nu}:=\lim_{\epsilon \to 0} \frac{\varphi^{U}[\gamma_{\mu\nu}]}{\epsilon^{2}} = \Tr{(w_{\lambda(0)}^{\dagger}w_{\lambda(0)}F_{\mu \nu}) }.
\end{equation*}\fi\indent
Notice that, while $F$ is gauge covariant, i.e. it transforms as $F\to U^{\dagger} F U$ under $w\to w U$, $\mathcal{U}_{\mu\nu}$ is a gauge invariant, i.e. it depends only on the infinitesimal path $\rho(t)$. In the gauge in which $w_{0}=\sqrt{\rho(0)}$, $\mathcal{U}_{\mu\nu}= \Tr{\left( \rho F_{\mu \nu}\right)}$ acquires the meaning of a \emph{mean Uhlmann curvature} (MUC).\\\indent
It can be shown that $\mathcal{U}_{\mu\nu} = \frac{i}{4}\Tr \rho[L_{\mu},L_{\nu}]$~(see Section I in~\footnote{See Supplemental Material}).
Such expression bears a striking resemblance with a pivotal quantity of quantum metrology, the Fisher information matrix (FIM), defined as $J_{\mu\nu}= \frac{1}{2}\Tr \rho\{L_{\mu},L_{\nu}\}$. The FIM determines a figure of merit of the estimation precision of parameters labelling a quantum state, known as the Cram\'er-Rao bound (CRB)~\cite{Helstrom1976,*Holevo2011,*Paris2009}. Given a set of locally unbiased estimators $\hat{\lambda}$ of the parameters $\lambda\in\mathcal{M}$, the covariance matrix $\Cov(\hat{\lambda})_{\mu\nu}=\langle (\hat{\lambda}_{\mu}-\lambda_{\mu})(\hat{\lambda}_{\nu}-\lambda_{\nu})\rangle$ is lower bounded (in a matrix sense) as follows
\ifnowc\begin{equation}\label{eq:CRB}
\Cov(\hat{\lambda})\ge J^{-1}.
\end{equation}\fi\indent
For single parameter estimation, the CRB can always be saturated by the projective measurement on the SLD eigenbasis. However, in a multi-parameter scenario this is not always the case, due to the non-commutativity of measurements associated to independent parameters. Within the framework of quantum local asymptotic normality (QLAN)~\cite{Hayashi2008,*Kahn2009,*Gill2013,*Yamagata2013}, one can prove that the multi-parameter CRB is attainable iff $\mathcal{U}_{\mu\nu}=0$ for all $\lambda_{\mu}$, $\lambda_{\nu}$~\cite{Ragy2016}. In this sense, $\mathcal{U}_{\mu\nu}$ marks the \emph{incompatibility} between $\lambda_{\mu}$ and $\lambda_{\nu}$, and such incompatibility arises from the inherent quantum nature of the underlying physical system. 
For a two-parameter model, the discrepancy between the \emph{attainable} multi-parameter bound and the CRB can be estimated by  the ratio $2|\mathcal{U}_{\mu\nu}|/{\Det{J}}$, and the MUC is upper bounded by (see Section V in~\cite{Note2})
\ifnowc\begin{equation}\label{UBound}
|\mathcal{U}_{\mu\nu}|\le\sqrt{\Det{J}}/2.
\end{equation}\fi\indent
When saturated, bound~(\ref{UBound}) marks the \emph{condition of maximal incompatibility}, in which the quantum indeterminacy in the estimation problem reaches the order of $\Det(J)^{-1/2}$, the same of the CRB~(\ref{eq:CRB}).\\\indent 
\emph{Dissipative quadratic models}.--
We now investigate the scaling law of the MUC, in dissipative Markovian models whose dynamics are generated by a master equation of Lindblad type~\cite{Breuer2002}
\ifnowc\begin{equation}
\frac{d\rho}{dt} =\mathcal{L}\rho= -i [\mathcal{H},\rho] + \sum_{\alpha}(2 \Lambda_{\alpha}\rho \Lambda_{\alpha}^{\dagger} -\{\Lambda_{\alpha}^{\dagger}\Lambda_{\alpha},\rho \}).  
\end{equation}\fi
The Hamiltonian is assumed quadratic in the fermion operators, i.e. $\mathcal{H}:=\bm{\omega}^{T} H \bm{\omega}$, where $\bm{\omega}:=(\omega_{1}\dots\omega_{2n})^{T}$ is a vector of Majorana operators: $\omega_{2k-1}:=c_{k}+c^{\dagger}_{k}$, $\omega_{2k}:=i(c_{k}-c_{k}^{\dagger})$, with $k=1\dots n,$ where $c_{k}$ and $c_{k}^{\dagger}$ are annihilation and creation operators. $H=-H^{T}$ is a $2n\times 2n$ Hermitian matrix. $\Lambda_{\alpha}=\bm{l}_{\alpha}^{T}\bm{\omega}$ are bath operators with $\bm{l}_{\alpha}:=(l_{1}^{\alpha},\dots,l_{2n}^{\alpha})^{T}\in\mathbb{C}^{2n}$.
\\\indent The Liouvillian $\mathcal{L}$ can be diagonalised exactly, and under certain conditions~\cite{Prosen2010a}, it admits a unique NESS $\rho$, which is Gaussian. A Gaussian state is completely specified by its correlation matrix $\Gamma_{jk}:=1/2\Tr{\rho[\omega_{j},\omega_{k}]}$. Let $\lambda\in\mathcal{M}$ be the set of parameters on which $H$ and $\bm{l}_{\alpha}$'s depend. Due to uniqueness, $\mathcal{M}$ parametrises the admissible NESS $\rho(\lambda)$. The correlation matrix of the NESS is the solution of the Lyapunov equation $X\Gamma + \Gamma X^{T} = Y$, with $X:=4[i H + \Re{(M)}]=X^{*}$, and $Y:=-i8\Im{(M)}=Y^{\dagger}=-Y^{T}$, where $M_{jk}:
=\sum_{\alpha} l^{\alpha}_{j}(l^{\alpha}_{k})^{*}=(M^{\dagger})_{jk}$ is called bath matrix.\\\indent 
According to~\cite{Prosen2010a}, the condition of NESS uniqueness is $\Delta:=2\min_{j}{\Re{(x_{j})}}\ge 0$, where $x_{j}$ is an eigenvalue of $X$, and $\Delta$ is the Liouvillian spectral gap. When this condition is met, any state will eventually decay into the NESS in a time scale $\tau\simeq1/\Delta$.
In the thermodynamical limit $n\to\infty$ a vanishing gap $\Delta(n)\to 0$ may be accompained, though not-necessarily, by non-differentiable properties of the NESS~\cite{Prosen2008,Znidaric2015}. For this reason, the scaling of $\Delta(n)$ has been used as an indication of NESS criticality~\cite{Prosen2010,*Znidaric2011,*Horstmann2013,*Cai2013,Znidaric2015}.  NESS-QPT has been investigated through the scaling of the Bures metrics~\cite{Zanardi2007a,CamposVenuti2007}, whose super-extensivity has been connected to a vanishing $\Delta$~\cite{Banchi2014}. 
%We show that a similar relation between the scaling properties of the dissipative gap and the MUC exists, namely 
%ifnowc\begin{equation*}
%\frac{| \mathcal{U}_{{\mu\nu}} |}{n} \le \frac{P_{\Gamma}}{\Delta^{2}}(||\partial_{\mu} Y ||_{\infty} +2 ||\partial_{\mu} X ||_{\infty})(||\partial_{\nu} Y ||_{\infty} +2 ||\partial_{\nu} X ||_{\infty})
%\end{equation*}\fi
 %where $P_{\Gamma}=||\Gamma\otimes\one+\one\otimes\Gamma/(\one+\Gamma^{\otimes 2})^{2}||_{\infty}$, and $||B||_{\infty}$ is the largest singular value of a matrix $B$. 
A similar relation between the super-extensivity of the MUC and $\Delta$ is implied by the inequality $||\mathcal{U}||_{\infty}\le||J||_{\infty}/2=||g||_{\infty}$ (see Section V in~\cite{Note2}), i.e. $\frac{| \mathcal{U}_{{\mu\nu}} |}{n} \le \frac{P_{\Gamma}}{\Delta^{2}}(||dY ||_{\infty} +2 ||dX ||_{\infty})^{2}$, where $||B||_{\infty}$ indicates the largest singular value of a matrix $B$, $P_{\Gamma}:=||(1+\Gamma\otimes\Gamma)^{-1}||_{\infty}$ and $g$ is the Bures metric tensor, which, except in pathological cases~\cite{Safranek2017}, is equal to $g=J/4$. This bound shows that if $P_{\Gamma}\simeq \mathcal{O}(1)$, a scaling of $|\mathcal{U}|\propto n^{\alpha+1}$ entails a dissipative gap that vanishes at least as $\Delta\propto n^{-{\alpha}/2}$, providing a relation between the dynamical properties of the NESS-QPT and the MUC.\\\indent
%
 %However, this relation bears a intriguing properties of $\mathcal{U}$ comes from its intrinsic quantum nature. Indeed, from the expression~(\ref{MUC}) it is clear that MUC arises from the commutator of two SLD, and, as such, its super-extensive properties cannot arise from the classical fluctuations, as in equilibrium thermal phase transition, but can only arise as a consequence of non-commutativity of close-by density matrices $\rho(\lambda)$ and $\rho(\lambda+d\lambda)$. In this sense, $\mathcal{U}$ is a signature of criticality associated to quantum fluctuations, as it cannot be sensitive to criticality induced by classical fluctuations, i.e. those associated only to changes in eigenvalues and not eigenstates of the density matrices.
\begin{figure}
	\includegraphics[width=\linewidth]{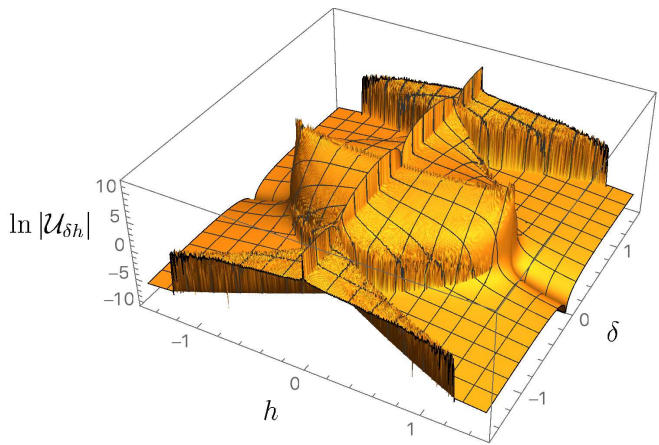}
    \caption{The MUC $|\mathcal{U}_{\delta h}|$ for the boudary driven XY model, for $n=300$. The qualitative behaviour of MUC maps the phase diagram quite faithfully. The discontinuity accross the critical line $h=h_{c}:=|1-\delta^{2}|$ signals the transition between LRMC and short range phases. $\kappa_{L}^{+}=0.3$, $\kappa_{L}^{-}=0.5$,$\kappa_{R}^{+}=0.1$, $\kappa_{R}^{-}=0.5$. The qualitative features remains unchanged for different values of $\kappa_{L,R}^{\pm}.$}\label{UDBXY}
\end{figure}
\begin{figure}
\includegraphics[width=\linewidth]{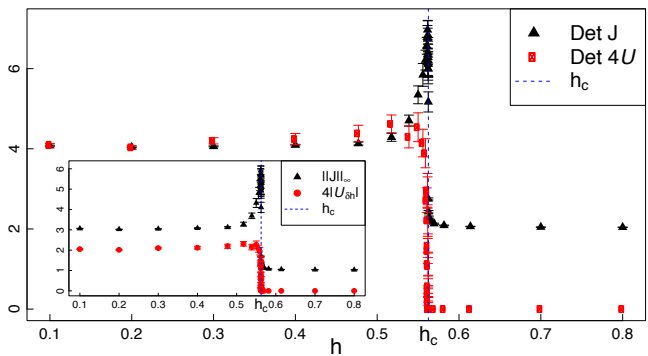}    
\caption{Boundary driven XY model. Scaling laws of the determinants (main) and maximal eigenvalues (inset) of the Fisher information matrix $J$ and mean Uhlmann curvature $\mathcal{U}$ for different values of $h$, with $\delta=1.25$ and $h_{c}=|1-\delta^{2}|$. The laws do not depend on the particular values of the $\kappa^{\pm}_{R,L}$. The scalings are the results of fits on numerical data, with size ranging in $n\in[20,2000]$.}\label{fig:Scaling}
\end{figure}
\begin{table}[htp]
\ifnowc
\begin{center}
\begin{tabular}{lccccc}
\hline
Phase&Parameters&$\Delta$  & $||J||_{\infty}$&\Det{J }&$|\mathcal{U}_{\delta h}|$\\
\hline
Critical &$ h=0$ & $n^{-3}$ & $n^{6}$ & $n^{7}$&$n^{3}$\\
Long range&$0<|h|<h_{c}$& $n^{-3}$& $n^{3}$&$n^{4}$ &$n^{2}$\\
Critical &$h\simeq h_{c}$& $n^{-5}$& $n^{6}$&$n^{7}$ &$n^{0}$\\
Short range&$h > h_{c}$& $n^{-3}$& $n$&$n^{2}$ &$n^{0}$\\
Critical&$\delta=0,|h|<h_{c}$& $n^{-3}$& $n^{2}$&$n^{8}$ &$n^{3}$\\
\hline
\end{tabular}
\end{center}
\fi
\caption{Here we show a comparison between the scaling laws for: the dissipative gap $\Delta$~\cite{Prosen2008}, the largest eigenvalue $||J||_{\infty}$ of the FIM~\cite{Banchi2014}, the determinant of $J$ and the largest eigenvalue $||\mathcal{U}||_{\infty}=|\mathcal{U}_{\delta h}|=\sqrt{\Det\,\mathcal{U}}$ of the MUC  for each phase of the boundary driven XY model~\cite{Prosen2008}.}\label{table}
\end{table}
Let's apply the above analysis to a specific model, the boundary-driven spin-1/2 XY chain~\cite{Prosen2008}. In this model, an open chain of spin-1/2 particles interacts via the $XY$-Hamiltonian, 
\ifnowc\begin{equation}\label{XY}
H_{XY}\!\!=\!\sum_{j=1}^{n-1}\! \left(\frac{1\!+\!\delta}{2}\sigma_j^x \sigma_{j+1}^x\!+\!\frac{1\!-\!\delta}{2}\sigma_j^y \sigma_{j+1}^y\right)\! +\!\sum_{j=1}^{n}\lambda\sigma^z_j,
\end{equation}\fi
where the $\sigma_{j}^{x,y,z}$ are Pauli operators acting on the spin on the $j$-th site.
At each boundary, the chain is in contact with two different reservoirs, described by Lindblad operators $\Lambda^{\pm}_{L}=\sqrt{\kappa_{L}^{\pm}}(\sigma_{j}^{x}\pm i\sigma_{j}^{y})/2$ and $\Lambda^{\pm}_{R}=\sqrt{\kappa_{R}^{\pm}}(\sigma_{j}^{x}\pm i\sigma_{j}^{y})/2$. A Jordan-Wigner transform converts the system into a quadratic fermionic dissipative model with Gaussian NESS~\cite{Prosen2008,Prosen2010}. The system experiences different phases as the anisotropy $\delta$ and magnetic field $h$ are varied. For $h<h_{c}:=|1-\delta^{2}|$ the chain exibits long-range magnetic correlations (LRMC) and high sensitivity to external parameter variations. For $h>h_{c}$ and along the lines $h=0$ and $\delta=0$ the model shows short-range correlations, with correlation function $C_{jk}:=\langle\sigma^{z}_{j}\sigma^{z}_{k}\rangle - \langle\sigma^{z}_{j}\rangle\langle\sigma^{z}_{k}\rangle$ exponentially decaying: $C_{jk}\propto\exp{-|j-k|/\xi}$, with $\xi^{-1}\simeq 4\sqrt{2(h-h_{c})/h_{c}}$. In both long and short range phases, the dissipative gap closes as $\Delta=\mathcal{O}(n^{-3})$ in the thermodynamical limit $n\to\infty$. The critical line $h=h_{c}$, is characterised by power-law decaying correlations $C_{jk}\propto|j-k|^{-4}$, and $\Delta=\mathcal{O}(n^{-5})$. Therefore, the scaling law of $\Delta$ cannot distinguish long and short range phases, and can only detect the actual critical line $h=h_{c}$. Likewise, $\Delta$ does not identify the transition from the LRMC phase to the $\delta=0$ and $h=0$ lines.\\\indent 
In table~\ref{table}, the MUC scaling law is compared with the scaling of $||J||_{\infty}$, $\Det{J}$ and $\Delta$ in each region of the phase diagram. Fig.~\ref{UDBXY} clearly shows that $|\mathcal{U}_{\delta h}|$ maps faithfully the phase diagram. A super-extensive behaviour of the MUC characterises the LRMC phase with a scaling $|\mathcal{U}_{\delta h}|=\mathcal{O}(n^{2})$, while in the short range phase the MUC is size independent. Thus, differently from $\Delta$, the MUC discriminates these phases, with no need of crossing the critical line $h=h_{c}$. Fig.~\ref{fig:Scaling} shows that in the LRMC phase, the scaling law of the MUC saturates the upper bound~(\ref{UBound}), in contrast to the short range phase. This shows the striking different nature of the two phases. In the LRMC region, the system behaves as an inherently two-parameter quantum estimation model, where the parameter incompatibility cannot be neglected even in the thermodynamical limit. On the short-range phase, instead, the system is asymptotically quasi-classical. The critical line $\delta=0$ (with $|h|\le h_{c}$) and the critical line $h=0$, which mark regions of short range correlations embedded in a LRMC phase, show a MUC which grows super-extensively, with scaling $\mathcal{O}(n^{3})$, and a nearly saturated inequality~(\ref{UBound}). In the critical line $h\simeq h_{c}$, despite the spectacular divergence of $||J||_{\infty}\simeq\mathcal{O}(n^{6})$, the scaling law of $|\mathcal{U}_{\delta h}|$ drops to a constant, revealing an asymptotic quasi-classical behaviour of the model at the phase transition. %According to ref~\cite{Banchi2014}, the critical line $\delta=0$ should show a scaling of $||g||_{\infty}$  as $\mathcal{O}(n^{2})$. However, it can be shown that $||g||_{\infty}$ is lower bounded by $||\mathcal{U}||_{\infty}$ (see appendix XXX). This  entails that the $\mathcal{O}(n^{2})$ behaviour of $||g||_{\infty}$ is correct only for relatively small values of $n$, whereas for larger $n$ values, a scaling faster than $\mathcal{O}(n^{3})$ is expected.\\
%%%%%A particularly remarkable feature appears at the critical line $h\simeq h_{c}$. The MUC shows a sudden transition between the $\mathcal{O}(n^{2})$ scaling of the LRMC phase to the $\mathcal{O}(n^{0})$ scaling of the short range phase. The multi-parameter statistical model switch from quantum to quasi-classical, i.e. $\delta$ and $h$ becomes compatible, as the critical line is crossed from the LRMC to the short-range phase. A
%%%%%, whereas $g$ scales as $\mathcal{O}(n^{6})$ in the close proximity of $h\simeq h_{c}$
%%%%%, where, in contrast to Bures metrics $g$, the MUC does not behave super-extensively.
%%%%%Such insensitivity of the MUC to this critical line signals that the multi-parameter statistical model is quasi-classical,   might be due to the nature of the dominant fluctuations. It is known that the geometric tensor is equivalent, up to a factor, to the Fisher information $=4 g$, and the latter can be decomposed as $F=F_{C}+F_{Q}$, into a classical and quantum contribution, $F_{C}$ and $F_{Q}$, respectively~\cite{Braunstein1994,Pezze2014,Hauke2015}. We find that the Fisher Information at the critical line $h\simeq h_{c}$ is dominated by the classical contributions, which scales as $\mathcal{O}(n^{6})$, while the quantum contribution scales as $\mathcal{O}(n^{5})$.
\\\indent
\emph{Translationally invariant systems.}--
We consider now a translationally invariant quadratic fermionic dissipative chain, with finite range interactions and local reservoirs. For infinite large systems, we analytically demonstrate the relation between the singularity of the MUC and the criticality of the model. To this end, following~\cite{Eisert2010,Honing2012}, we define criticality by the divergence of correlation length $\xi^{-1} := - \lim_{|r|\to \infty} \frac{\ln {||\gamma(r)||}}{|r|}$, where $[\gamma(r-s)]_{\beta \beta'}:=1/2\Tr{\rho[\omega_{r,\beta},\omega_{s,\beta'}]}$ is the spacial correlation function, and $\omega_{r,\beta}$ labels a Majorana fermion of either of the two species $(\beta=1,2)$ on the $r$-th site. In the thermodynamical limit, the covariance matrix of the NESS can be represented by its Fourier transform, i.e. $\gamma(r)=\frac{1}{2\pi}\int_{0}^{2\pi}\tilde{\gamma}{(\phi)} e^{-i\phi r}$. In Fourier components, the Lyapunov equation becomes a $2\times2$ matrix equation
\ifnowc\begin{equation}\label{CLEphi}
	\tilde{x}(\phi) \tilde{\gamma}(\phi)+\tilde{\gamma}(\phi)\tilde{x}^{T} (-\phi)= \tilde{y}(\phi),
\end{equation}\fi
 where $\tilde{x}(\phi)$ and $\tilde{y}(\phi)$ are obtained from the Fourier transform of $X$ and $Y$~\cite{Eisert2010,Honing2012}.  Through the analytical extension of $\tilde{\gamma}(z):= \tilde{\gamma}(e^{i\phi}\to z)$ in the complex plane, the spatial correlations can be expressed as $ \gamma(r) =\sum_{\bar{z}\in S_{1}} \text{Res}_{\bar{z}}[z^{r-1}\tilde{\gamma}{(z)}]$,
 where $ \text{Res}_{\bar{z}}$ indicates the residues inside the unit circle $S_{1}:|z|\le1$. Since $\tilde{\gamma}(z)$ has at most simple poles, criticalities can occur only when isolated poles approach $S_{1}$ from inside~\cite{Eisert2010,Honing2012}. \\
Thanks to the convolution theorem, the MUC \emph{per site} can be expressed as
\ifnowc\begin{equation}\label{UPhi}
\bar{\mathcal{U}}_{\mu\nu}:=\lim_{n\to\infty} \frac{\mathcal{U}_{\mu\nu}}{n}=\frac{1}{(2\pi)}\int_{-\pi}^{\pi}d\phi  \,\,u_{\mu\nu}(\phi),
\end{equation}\fi
where
\ifnowc\begin{equation*}\label{uuphi}
u_{\mu\nu}(\phi) =\left\{ \begin{array}{ll }
      \frac{i}{4}\frac{\Tr\{\tilde{\gamma}{(\vv{\phi})}[\partial_{\mu}\tilde{\gamma}{(\vv{\phi})},\partial_{\nu}\tilde{\gamma}{(\vv{\phi})}]\}}{(1-\Det{\tilde{\gamma}{(\vv{\phi})}})^{2}}& \Det{\tilde{\gamma}{(\vv{\phi})}}\neq 1\\
       0 & \Det{\tilde{\gamma}{(\vv{\phi})}}=1
\end{array} \right..
\end{equation*}\fi\indent
We show in section III of~\cite{Note2} that a singularity of $\bar{\mathcal{U}}$ signals the occurrence of a criticality. Specifically, employing the analytical extension in the complex plane of $u_{\mu\nu}{(\phi)}$ leads to
\ifnowc\begin{equation}\label{eq:Uz}
\bar{\mathcal{U}}_{\mu\nu}=\sum_{\bar{z}'\in S_{1}} \text{Res}_{\bar{z}'}[z^{-1} u_{\mu\nu}(z)].
\end{equation}\fi\indent 
Notice that $u_{\mu\nu}(z)$ has at most isolated poles, due to its rational dependence on $z$.
Assume that as $\lambda\to\lambda_{0}\in\mathcal{M}$, a pole $\bar{z}_{0}$ of $u_{\mu\nu}(z)$ approaches $S_{1}$,  which is the only condition under which $\bar{\mathcal{U}}$ is singular in $\lambda_{0}$.
As this happens, a pole $\bar{z}$ of $\tilde{\gamma}(z)$ approaches $\bar{z}_{0}$, determining a diverging correlation length (see Section III in~\cite{Note2}). Therefore the singular behaviour of the Uhlmann phase represents a sufficient criterion for criticality in such systems. Notice that such criticalities are necessarily accompanied by the closure of the dissipative gap, however, the converse is in general not true (see Section IV in~\cite{Note2}).
Indeed, a singularity in the MUC may only arise as the result of criticality and are otherwise insensitive to a vanishing dissipative gaps.\\\indent
\begin{figure}[t]
\includegraphics[width=\linewidth]{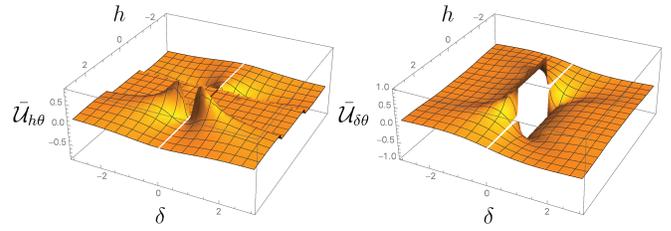}
\caption{The mean Uhlmann curvature per number of sites $\bar{\mathcal{U}}$ for the rotated XY model with local reservoirs. The dependence of $\bar{\mathcal{U}}_{h\theta}$ (left) and of $\bar{\mathcal{U}}_{\delta\theta}$ (right) on the parameters $\delta$ e $h$. The mean Uhlmann curvature shows a singular behaviour in the critical regions of the model. $\mathcal{U}_{h\theta}$ is discontinuous in the $XY$ critical points $|h|=1$, and $\mathcal{U}_{\delta\theta}$ is discontinuous in the $XX$ type criticalities $\delta=0$, $h< 1$. }\label{UXY}
\end{figure}
These features are exemplified in the following translational invariant dissipative fermionic chain: the rotated XY model with periodic boundary conditions~\cite{Carollo2005,*Pachos2006d},
$H=R(\theta)H_{XY}R(\theta)^{\dagger}$, with $R(\theta)=e^{-i\frac{\theta}{2}\sum_{j}\sigma_{j}^{z}}$ and 
\ifnowc\begin{equation}\label{RotatedXY}
H_{XY}\!=\!\sum_{j=1}^{n} \!\left(\frac{1\!+\!\delta}{2}\sigma_j^x\,\, \sigma_{j+1}^x\!+\!\frac{1\!-\!\delta}{2}\sigma_j^y \sigma_{j+1}^y \!+\!\lambda\sigma^z_j\right),
\end{equation}\fi
where each site $j$ is coupled to two local reservoirs with Lindblad operators $\Lambda_{j}^{\pm}=\epsilon \mu \sigma_{j}^{\pm}$. The spin-system is converted into a quadratic fermionic model via Jordan-Wigner transformations. The Liouvillian spectrum can be solved exactly~\cite{Prosen2008,Prosen2010a,Horstmann2013} and it is independent of $\theta$. In the weak coupling limit $\epsilon\to 0$, the symbol function of the NESS correlation matrix reads $\tilde{\gamma}(\phi)=\bm{\gamma}^{T}\cdot\bm{\sigma}$, where $\bm{\sigma}:=(\sigma_{x},\sigma_{y},\sigma_{z})^{T}$, and $\bm{\gamma}= g [t(\phi) \cos\theta,-1, t(\phi) \sin\theta]^{T}$,
with $g= \frac{\nu^{2}-\mu^{2}}{\nu^{2}+\mu^{2}} \frac{1}{1+t(\phi)^{2}}$ and $t(\phi) :=\delta \sin\phi/(\cos\phi -h)$. The system shows criticality in the same critical regions of the $XY$ hamiltonian model~\cite{Horstmann2013}. 
By using expression~(\ref{eq:Uz}) we can calculate the exact values of the mean Uhlmann curvature. We find that $\bar{\mathcal{U}}_{\delta h}$ vanishes, while $\bar{\mathcal{U}}_{\delta\theta}$ and $\bar{\mathcal{U}}_{h\theta}$ are plotted in Fig.~\ref{UXY}. As predicted, the Uhlmann curvature shows a singular behaviour only across criticality. In particular, $\mathcal{U}_{h\theta}$ is discontinuous in the $XY$ critical points $|h|=1$, while $\mathcal{U}_{\delta\theta}$ is discontinuous in the $XX$ type criticalities $\delta=0$, $h< 1$.\\\indent
\emph{Conclusions} --
We have introduced the mean Uhlmann curvature, and derived a formalism that emphasised its deep relation with estimation theory. Estimation theory and geometrical interpretations of the MUC together provide a crucial insight into the critical features of out-of-equilibrium quantum systems. The geometrical interpretation in terms of curvature provides an intuitive explanation as to why singularities of MUC emerge in criticalities, and leads to a unified interpretation for equilibrium and out-of-equilibrium QPT. In quantum metrology, the MUC accounts for the discrepancy between an inherently quantum and a quasi-classical multi-parameter estimation problem, sheding a new light onto the nature of correlations in NESS-QPT. We have explored the properties of the MUC in the physically relevant class of dissipative NESS-QPT exhibited by quadratic fermionic Liouvillian models. A relation between the super-extensive behaviour of the MUC and the criticality in the thermodynamical limit has been drawn. We studied the thermodynamical limit of translationally invariant systems and analytically demonstrated that a singular behaviour of the MUC implies criticality in the sense of diverging correlation length. We have employed specific prototypical models, showing that the scaling laws and the singularities of $\mathcal{U}$ map faithfully their phase diagrams. This approach goes well beyond the application to the important class of quadratic dissipative models analysed here, and introduces a tool suitable for the systematic investigation of out-of-equilibrium quantum critical phenomena. 

\ifnowc
We acknowledge fruitful discussions with Toma\v{z} Prosen. This work was partially supported by the Ministry of Education and Research of Italian Government.\fi
\ifnowc 
\bibliography{\bibliopath library} 
\else
\nobibliography{\bibliopath library} 
\fi
\clearpage
\widetext
\begin{center}
\textbf{\large Supplemental Materials: \Title}
\end{center}

\setcounter{equation}{0}
\setcounter{figure}{0}
\setcounter{table}{0}
\setcounter{page}{1}
\makeatletter
\renewcommand{\theequation}{S\arabic{equation}}
\renewcommand{\thefigure}{S\arabic{figure}}
\renewcommand{\bibnumfmt}[1]{[S#1]}
\renewcommand{\citenumfont}[1]{S#1}

\section{The mean Uhlmann curvature}\label{sec:MUC}
Here we will briefly review the idea of the Uhlmann geometric phase, and derive the expression of the mean Uhlmann curvature as a function of the symmetric logarithmic derivatives (SLDs). 
Given a density operator $\rho$ acting on a Hilbert space $\HH$ of dimension $n$, an exteded Hilbert space is defined by attaching an ancilla $a$: $\HH_{\text{ext}} = \HH\otimes\HH^{a}$. A purification is defined as any pure state $\psi\in\HH$ such that $\rho=\Tr_{a}{\ket{\psi}\bra{\psi}}$, where $\Tr_{a}$ is the partial trace over the ancilla. A standard choice for $\HH^{a}$ is the dual of $\HH$, then  $\HH_{\text{ext}}$ becomes the space of operator $w$ over $\HH$, with Hilbert-Schmidt scalar product $(w,v):=\Tr{(w^{\dagger}v)}$. 
Hence, a purification can be equivalently expressed in terms of any Hilbert-Schmidt operators $w$, called \emph{amplitudes}, such that 
\begin{equation}\label{rho}
\rho=w w^{\dag}
\end{equation}
The above equation, leaves a gauge freedom $U(n)$ in the choice of $w$, as any $w'=w U$ is an amplitude of the same $\rho$. Indeed, from the polar decomposition theorem we can always uniquely parametrise an amplitude as $w=\sqrt{\rho}U$. 

Given a pure state $\psi$, a similar $U(1)$ gauge freedom is obtained by the simple observation that any $\psi=e^{i\varphi} \psi'$ represents the same element of the projective Hilbert space. Let $\ket{\psi_{\lambda}}\bra{\psi_{\lambda}}$ be a family of pure states parameterised by $\lambda\in\mathcal{M}$, and let $\gamma:=\{\lambda(t)\in\mathcal{M},\, t\in[0,T]\}$ be a smooth closed path in the parameter manyfold $\mathcal{M}$. Given such a family we can choose any representative trajectory $\psi_{\lambda(t)}=e^{i\varphi(t)} \psi'_{\lambda(t)}$ in the Hilbert space.  If the trajectory chosen fulfils the prescription of parallel transport, i.e. $\bk{\psi_{\lambda(t)}| \frac{d}{dt}| \psi_{\lambda(t)}}=0$, then the phase difference $\varphi^{B}$ between initial and final state $\ket{\psi_{\lambda(T)}}=e^{i\varphi^{B}}\ket{\psi_{\lambda(0)}}$ is purely geometric in nature, i.e. it solely depends on the path $\gamma$, regardless of parameterisation and re-gauging. This phase is called Berry phase and its value reads $\varphi^{B}=\oint_{\gamma} A^{B}$, where $A^{B}:= \sum_{\mu} A^{B}_{\mu}d\lambda_{\mu}$ is the Berry connection one-form, whose components are $A^{B}_{\mu}:=i \bra{\psi_{\lambda}}\partial_{\mu}\ket{\psi_{\lambda}}$, where $\partial_{\mu}:=\partial/\partial \lambda_{\mu}$. By exploiting the Stokes theorem, we can convert the loop integral of $A^{B}$ to an integral $\varphi^{B}=\int_{S} F^{B}$ over a surface $S$ bounded by the path $\gamma$, where $ F^{B}:=dA^{B}=\frac{1}{2}\sum_{\mu \nu} F^{B}_{\mu\nu}d\lambda_{\mu}\wedge d\lambda_{\nu}$ is the Berry curvature two-form, whose components are $F^{B}_{\mu\nu}:=\partial_\mu A^{B}_\nu -\partial_\nu A^{B}_\mu$. The parallel transport condition is equivalent to choose the representative path $\psi_{\lambda(t)}$ that minimizes the length of the path on the Hilbert space measure by $l=\int_{0}^{T} d\tau \sqrt{\bk{\dot{\psi(\tau)}|\dot{\psi(\tau)}}}$. 

Similarly, we can have a smooth closed trajectory of density matrices, $\rho_{\lambda(t)}$, parametrised by a path $\gamma:$ $\lambda(t) \in \mathcal{M}$, $t\in[0,T]$, and, correspondingly, a path of Hilbert-Schmidt operators $w_{\lambda(t)}$ in $\HH_{\text{ext}}$. The choice of amplitudes is quite redundant due to the local $U(n)$ gauge freedom. Similarly to the pure state case, this redundancy can be mitigated by imposing the so called Uhlmann parallel transport condition, which prescribes that, given any two $\rho_{1}$ and $\rho_{2}$, their respective amplitudes $w_{1}$ and $w_{2}$ are parallel whenever 
\begin{equation}\label{dparcond}
w_{1}^{\dagger}w_{2}=w_{2}^{\dagger}w_{1}\ge 0.
\end{equation}
This is equivalently means that the chosen $w_{1}$ and $w_{2}$ are those that maximise their Hilbert Schmidt scalar product $(w_{1},w_{2}) := \Tr{(w_{1}^{\dagger} w_{2})}$, i.e. 
\[
	(w_{1},w_{2})= \max_{w'_{2}} |(w_{1},w'_{2})|:=\mathcal{F}(\rho_{1},\rho_{2})
\]
where the maximum is taken over all $w'_{2}$ purifying $\rho_{2}$. The above maximal value depends on $\rho_{1}$ and $\rho_{2}$ only, and it is equal to $\mathcal{F}(\rho_{1},\rho_{2})=\Tr{\sqrt{\sqrt{\rho_{1}}\rho_{2}\sqrt{\rho_{1}}}} $, the so called Uhlmann fidelity of $\rho_{1}$ and $\rho_{2}$.  Through the fidelity one can define a geometric measure of statistical indistinguishability between states $\rho_{1}$ and $\rho_{2}$~\cite{Braunstein1994}, the Bures distance:
\[
d^{2}_{B}(\rho_{1},\rho_{2}):=2(1-\mathcal{F}(\rho_{1},\rho_{2})).
\]
which, for infinitesimally closed states, defines a Riemannian metrics on the manifold of density operators, the Bures metrics:
\[
\sum_{\mu\nu} g^{\mu\nu} d\lambda_{\mu} d\lambda_{\mu} := d^{2}_{B} (\rho_{\lambda},\rho_{\lambda+d\lambda}). 
\]
Applied to any two neighbouring points $w_{\lambda(t)}$ and $w_{\lambda(t+dt)}$ of a smooth path of amplitudes, the parallel transport condition~(\ref{dparcond}) becomes:
\begin{equation}\label{parcond}
w^{\dagger} \dot{w}- \dot{w}^{\dagger} w = 0,
\end{equation}
where dots denote derivatives with respect to $t$. The maximisation of the overlap $(w_{\lambda(t)},w_{\lambda(t+dt)})$ is equivalent to the minimisation of the  ``velocity'' $v:=\sqrt{(\dot{w},\dot{w})}$, which in turns means that the path of amplitudes fullfilling the Uhlmann condition are those with the shortest length measured by $l:=\int_{0}^{T} d\tau \sqrt{(\dot{w}_{\lambda(\tau)},\dot{w}_{\lambda(\tau)} )}$.

According to~\cite{Uhlmann1989,*Dabrowski1989,*Dabrowski1990}, the parallel transport condition~(\ref{parcond}) is fullfilled by the following ansatz
\begin{equation}\label{ansatz1} 
\dot{w} = \frac{1}{2} L_{t} w,\qquad L_{t}^{\dagger}=L_{t}.
\end{equation}
$L_{t}$ can be determined by differentiating $\rho=w w^{\dagger}$ and inserting~(\ref{ansatz1}), which yields:
 \begin{equation}\label{SLD}
 \dot{\rho}=\frac{1}{2}\{L_{t},\rho\},
 \end{equation}
where $\{.,.\}$ is the anticommutator. $L_{t}$, known as the symmetric logarithmic derivative (SLD), is implicitly defined as the (unique) operator solution of~(\ref{SLD}) with the auxiliary requirement that $\bk{\psi | L _{t}| \psi} = 0$, whenever $\rho\ket{\psi}=0$. As mentioned in the main text, as far as the definition of the SLD is concerned, we will actually confine ourselves to full-rank density matrices. In the case of singular density matrices, quantities of interest to us can be calculated \emph{consistently} by a limiting procedure from the set of full rank matrices.  
In terms of $L_{t}$, the ``velocity'' can be cast as $v=\sqrt{\Tr{(\dot{w} \dot{w}^{\dagger})}}=1/2\sqrt{\Tr[L_{t}\rho L_{t}]}$, which in turn means that the Bures metrics can be expressed in the following form:
\begin{equation}\label{Bures}
	g_{\mu\nu}=\frac{1}{8}\Tr{(\rho\{L_{\mu},L_{\nu}\})}
\end{equation}
where $L_{\mu}$ is the restriction of $L_{t}$ along the coordinate $\lambda_{\mu}$, and it is determined by the analog of equation~(\ref{SLD}), $\partial_{\mu}\rho=\frac{1}{2}\{L_{\mu},\rho\}$, where ($\partial_{\mu}:=\partial/{\partial \lambda_{\mu}}$).
We can also define the operator-valued differential one-form $L:=\sum_{\mu}L_{\mu} d\lambda_{\mu}$. 
In the closed path  $\rho_{\lambda(t)}$, initial and final amplitudes are related by a unitary transformation, i.e. $w_{\lambda(T)}=w_{\lambda(0)}V_{\gamma}$. If the path of amplitudes $w_{\lambda(t)}$ fullfills the Uhlmann condition, $V_{\gamma}$ is a \emph{holonomy}, the non-Abelian generalisation of Berry phase~\cite{Uhlmann1986}. The holonomy is expressed as $V_{\gamma}=\mathcal{P}e^{i\oint_{\gamma} A}$, where $\mathcal{P}$ is the path ordering operator and $A$ is the Uhlmann connection one-form. 
The Uhlmann connection can be derived from the following ansatz~\cite{Uhlmann1989,*Dittmann1999}
\begin{equation}\label{ansatz2}
d w +i w A = \frac{1}{2} L w 
\end{equation}
which is the generalisation of~(\ref{ansatz1}) when the parallel transport condition is lifted. By differentiating $\rho=w w^{\dagger}$ and using the defining property of the SLD (see eq.~(\ref{SLD})), it follows that$A$ is Hermitian and it is implicitly defined by the equation
\[
A w^{\dagger} w + w^{\dagger} w A =i(w^{\dagger} dw- dw^{\dagger} w ),
\]  
with the auxiliary constraint that $\bk{\psi' | A | \psi'} = 0$, for $w\ket{\psi'}=0$. From eq.~(\ref{ansatz2}), it can be  checked that $A$ obeys the expected transformation rule of non-Abelian gauge potentials, $A\to U^{\dagger}_{t} A U_{t}+i U_{t}^{\dagger}dU_{t}$ under $w_{t}\to w_{t}U_{t}$, and that $L$ is gauge invariant. 

The analog of the Berry curvature, the Uhlmann curvature two-form, is defined as $F:=dA-i A\wedge A= \frac{1}{2} \sum_{\mu\nu} F_{\mu\nu}d\lambda_{\mu}\wedge d\lambda_{\nu}$. Its components $F_{\mu\nu}=\partial_\mu A_\nu -\partial_\nu A_\mu-i\left[A_{\mu},A_{\nu}\right]$ can be understood in terms of the Uhlmann holonomy per unit area associated to an infinitesimal loop in the parameter space. Indeed, for an infinitesimal parallelogram $\gamma_{\mu \nu}$, spanned by two independent directions $\hat{e}_{\mu}\delta_{\mu}$ and $\hat{e}_{\nu}\delta_{\nu}$ in the manifold, it reads
\[
 F_{\mu\nu}=\lim_{\delta \to 0} i \frac{1-V_{\gamma_{\mu,\nu}}}{\delta_{\mu}\delta_{\nu}},
\]
where $\delta \to 0$ is a shorthand of $(\delta_{\mu},\delta_{\nu})\to (0,0)$.

As already mentioned in the main text, the Uhlmann geometric phase is defined as
\begin{equation}\label{Uphi}
\varphi^{U}[\gamma]:=\arg{(w_{\lambda(0)},w_{\lambda(T)})}=\arg{\Tr{\left[w_{\lambda(0)}^{\dagger}w_{\lambda(T)}\right]}},
\end{equation}
and the Uhlmann phase per unit area for an infinitesimal loop reads
\[
 \mathcal{U}_{\mu\nu}:=\lim_{\delta \to 0} \frac{\varphi^{U}[\gamma_{\mu\nu}]}{\delta_{\mu}\delta_{\nu}} = \Tr{\left[w_{\lambda(0)}^{\dagger}w_{\lambda(0)}F_{\mu \nu}\right] }.
\]
We called the latter \emph{mean Uhlmann curvature} (MUC), on account of the expression $\mathcal{U}_{\mu\nu}= \Tr{\left( \rho F_{\mu \nu}\right)}=\langle F_{\mu\nu}\rangle$  that $\mathcal{U}$ takes in the special gauge $w_{0}=\sqrt{\rho(0)}$.

By taking the external derivative of the expression~(\ref{ansatz2}) and by using the property $d^{2}=0$, it can be shown that~\cite{Uhlmann1991}
\begin{eqnarray}
w F &= \frac{i}{4}\left(L\wedge L-2dL\right) w,\\
F w^{\dagger} &= \frac{i}{4}w^{\dagger}\left(L\wedge L+2dL\right).
\end{eqnarray} 
Multiplying the above expressions by $w^{\dagger}$ and $w$, respectively, and taking the trace yields
\begin{equation}\label{MUC}
\mathcal{U} = \Tr(w Fw^{\dagger}) = \frac{i}{4}\Tr(\rho L\wedge L),
\end{equation} 
where $\mathcal{U}:=1/2 \sum_{\mu\nu}\mathcal{U}_{\mu\nu}d\lambda_{\mu}\wedge d\lambda_{\nu}$ is a real-valued two-form whose components are $\mathcal{U}_{\mu\nu}=\frac{i}{4}\Tr(\rho [L_{\mu},L_{\nu}])$. 
The expressions of~(\ref{Bures}) and~(\ref{MUC}) reveal the common mathematical structure of MUC and metric tensor, which can be merged into a Hermitian matrix 
\begin{equation}\label{QFT}
I_{\mu \nu}:=\Tr{(\rho L_{\mu}L_{\nu})},
\end{equation}
called the quantum Fisher tensor (QFT)~\cite{Ercolessi2012,*Ercolessi2013,*Contreras2016}, such that  $g_{\mu\nu}=\Re{(I_{\mu\nu}})/4$ and $\mathcal{U}_{\mu\nu}=-\Im{(I_{\mu\nu})}/2$. 

 \section{Fermionic Gaussian states}\label{sec:Gauss}
 We will specialize our considerations to the case of systems described by fermionic Gaussian states. The fermionic Gaussian states are defined as density matrices $\rho$ that can be expressed as
 \begin{equation}\label{GS}
\rho=:e^{-\frac{i}{4}\bm{\omega}^{T} \Omega \bm{\omega}}/Z,\qquad Z:=\Tr{e^{-\frac{i}{4}\bm{\omega}^{T} \Omega \bm{\omega}}}. 
 \end{equation}
 
Here $\Omega$ is a $2n\times 2n$ real antisymmetric matrix, and $\bm{\omega}:=(\omega_{1}\dots\omega_{2n})^{T}$ is a vector of $2n$ Majorana fermion operators, defined as: $\omega_{2k-1}:=c_{k}+c^{\dagger}_{k}$, $\omega_{2k}:=i(c_{k}-c_{k}^{\dagger})$, with $k=1\dots n,$ where $c_{k}$ and $c_{k}^{\dagger}$ are annihilation and creation operators of standard fermions, respectively. The anticommutation relations of the Majorana fermion operators read $\{\omega_{j},\omega_{k}\}=2\delta_{jk}$. The Gaussian state is completely specified by the two-point correlation matrix $\Gamma_{jk}:=1/2\Tr{(\rho[\omega_{j},\omega_{k}])}$, which is an imaginary antisymmetric matrix. One can show that $\Gamma$ and $\Omega$ can be simultaneously cast in a canonical form by an orthogonal matrix $Q$
\[
\Gamma=Q \bigoplus_{k=1}^{n}\left(\begin{array}{cc}0 & i\gamma_k \\- i\gamma_k & 0\end{array}\right)Q^{T},\quad \Omega=Q \bigoplus_{k=1}^{n}\left(\begin{array}{cc}0 & \Omega_{k} \\- \Omega_k & 0\end{array}\right)Q^{T},
\]

and their eigenvalues are related by $\gamma_{j}=\tanh{(\Omega_{j}/2)}$, which implies that$|\gamma_{j}|\le1$. Correspondingly, the density matrix can be factorised as
\[
\rho=\bigotimes_{k=1}^{n}\frac{\one+i\gamma_{k} z_{2k-1}z_{2k}}{2}, 
\]

where $\bm{z}=(z_{1},\dots,z_{2n})^{T}:=Q\bm{\omega}$ are the Majorana fermions in the eigenmode representation. Notice that $|\gamma_{k}|=1$ corresponds to the fermionic mode $\tilde{c}_{k}=1/2(z_{2k-1}+z_{2k})$ being in a pure state.

For a Gaussian fermionic state, all odd-order correlation functions are zero, and all even-order correlations, higher than two, can be obtained from  $\Gamma$ by WickÕs theorem~\cite{Bach1994} , i.e. $\Tr(\rho \omega_{k_{1}} \omega_{k_{2}}...\omega_{k_{2p}})=\Pf(\Gamma_{k_{1}k_{2}\dots k_{2p}})$,  where $1 \le k_{1} < . . . < k_{2p} \le 2n$ and $\Gamma_{k_{1}k_{2}\dots k_{2p}}$ is the corresponding $2p \times 2p$ submatrix of $\Gamma$. $\Pf(\Gamma_{k_{1}k_{2}\dots k_{2p}})^{2} = \det\Pf(\Gamma_{k_{1}k_{2}\dots k_{2p}})$ is the Pfaffian. An especially useful case is the four-point correlation function
\begin{equation}\label{Pf}
	\Tr{(\rho\omega_{j} \omega_{k}\omega_{l}\omega_{m})}= a_{jk}a_{lm}-a_{jl}a_{km}+a_{jm}a_{kl},
\end{equation}    
where $a_{jk}:=\Gamma_{jk}+\delta_{jk}$.

We would like to derive a convenient expression for the QFT for Gaussian fermionic states. In order to do this, we first derive the SLD in terms of correlation matrix $\Gamma$. Due to the quadratic dependence of~(\ref{GS}) in $\bm{\omega}$, and following the arguments of~\cite{Monras2013,*Jiang2014}, it can be shown that $L$ is a quadratic polynomial in the Majorana fermions
 \begin{equation}\label{SLDGS}
 L_{}=: \frac{1}{2}\bm{\omega}^{T}\cdot K \bm{\omega} + \bm{\zeta}_{}^{T}\bm{\omega} +\eta_{},
 \end{equation}
 where $K=\sum_{\mu}K_{\mu}d\lambda_{\mu}$, with $K_{\mu}$ a $2n\times2n$ hermitian antisymmetric matrix, $\bm{\zeta}=\bm{\zeta}_{\mu} d\lambda_{\mu}$, with $\bm{\zeta}_{\mu}$ a $2n$ real vector, and $\eta=\eta_{\mu}d\lambda_{\mu}$ a real valued one-form. From the property that $\Tr{(\rho \omega_{k})}=0$ for any $1\le k \le 2n$, it is straightforward to show that the linear term in~(\ref{SLDGS}) is identically zero
 \[
 0=\Tr{(\omega_{k} d \rho )}= \frac{1}{2}\Tr(\omega_{k}\{L,\rho\}) = \frac{1}{2}\Tr(\rho\{\bm{\zeta}^{T}\bm{\omega},\omega_{k}\})=\zeta^{k}. 
 \]
where $\zeta^{k}$ is the $k$-th component of $\bm{\zeta}$, and in the third equality we took into account that the odd order correlations vanish. The quantity $\eta$ can be determined from the trace preserving condition $\Tr{(d{\rho})}=\Tr{(\rho L)}=0$
\begin{equation}\label{eta}
\eta=-\frac{1}{2}\Tr{(\rho \bm{\omega}^{T}K \bm{\omega})}=\frac{1}{2}\Tr{(K_{} \cdot \Gamma)}.
\end{equation}
 In order to determine $K_{}$, we take the differential of $\Gamma_{jk}=1/2\Tr{(\rho[\omega_{j},\omega_{k}])}$
 \begin{align}
 d\Gamma_{jk}=\frac{1}{2}\Tr{(d \rho[\omega_{j},\omega_{k}])}&=\frac{1}{4}\Tr{(\{\rho,L_{}\}[\omega_{j},\omega_{k}])}\nonumber\\
 &= \frac{1}{8}\Tr{(\{\rho,\bm{\omega}^{T}K_{}\bm{\omega}\}[\omega_{j},\omega_{k}])}+ \eta_{}\frac{1}{2}\Tr{(\rho[\omega_{j},\omega_{k}])}\nonumber\\
 &=\frac{1}{16}\sum_{lm}K_{}^{lm}\Tr{(\rho\{[\omega_{l},\omega_{m}][\omega_{j},\omega_{k}]\})}+\eta_{} \Gamma_{jk}\nonumber\\
 &=(\Gamma K_{} \Gamma-K_{})_{jk} + \left[\eta - \frac{1}{2}\Tr{(K_{}\cdot \Gamma)}\right]\Gamma_{jk},
 \end{align}
 where the last equality is obtained with the help of eq.~(\ref{Pf}) and using the antisymmetry of $\Gamma$ and $K_{}$. Finally, according to eq.~(\ref{eta}), the last term vanishes and we obtain the following (discrete time) Lyapunov equation
\begin{equation}\label{DLE} 
d\Gamma = \Gamma K_{}\Gamma -K_{}.
\end{equation}
The above equation can be formally solved by
\[
K_{}=-(\one-\Ad_{\Gamma})^{-1}(d \Gamma),
\]
where $\Ad_{\Gamma}(X):=\Gamma X \Gamma^{\dagger}$ is the adjoint action. In the eigenbasis of $\Gamma$ it reads
\begin{equation}\label{Kappa}
(K_{})_{jk}=-\frac{(d\Gamma)_{jk}}{1-\gamma_{j}\gamma_{k}}= -\frac{d\Omega_{k}}{2}\delta_{jk}+\tanh{\frac{\Omega_{j}-\Omega_{k}}{2}} \bk{j|dk},
\end{equation}
where, in the second equality, we made use of the relation $\gamma_{k}=\tanh{(\Omega_{k}/2)}$, which yields the following diagonal $(d\Gamma)_{jj}=(1-\gamma_{j}^{2})d\Omega_{j}$ and off-diagonal terms $(d\Gamma)_{jk}=(\gamma_{k}-\gamma_{j})\bk{j| dk}$. This expression is well defined everywhere except for $\gamma_{j}=\gamma_{k}=\pm 1$, where the Gaussian state $\rho$ becomes singular (i.e. it is not full rank). In this condition, the expression~(\ref{Kappa}) for the SLDs may become singular. Nevertheless,  the boundness of the function $|\tanh{\frac{\Omega_{j}-\Omega_{k}}{2}}|\le 1$ in~(\ref{Kappa}) shows that such a singularity is relatively benign. Thanks to this, we can show that the condition $\gamma_{j}=\gamma_{k}=\pm 1$ produces, at most, removable singularities in the QFT (cf.~\cite{Safranek2017}). This allows the QFT to be extended by continuity from the set of full-rank density matrices to the submanifolds with $\gamma_{j}=\gamma_{k}=\pm 1$.

Knowing the expression for the SLDs, we can calculate the QFT by plugging $L_{\mu}=\frac{1}{2}[\bm{\omega}^{T}K_{\mu} \bm{\omega} - \Tr{(K_{\mu}\cdot \Gamma)]}$ into $I_{\mu \nu}:=\Tr{(\rho L_{\nu}L_{\mu})}$. Making use of~(\ref{Pf}) and exploiting the antisymmetry of both $\Gamma$ and $K$ leads to
\begin{equation}\label{eq:QFT}
I_{\mu\nu}=\frac{1}{2}\Tr{[(\one+\Gamma)K_{\mu}(\one-\Gamma)K_{\nu}]}
=\frac{1}{2}\sum_{jk}(1+\gamma_{j})(1-\gamma_{k})K_{jk}K_{kj}
=\frac{1}{2}\sum_{jk}\frac{(1+\gamma_{j})(1-\gamma_{k})}{(1-\gamma_{j}\gamma_{k})^{2}}(\partial_{\mu}\Gamma)_{jk}(\partial_{\nu}\Gamma)_{kj},
\end{equation}
where the last equality is obtained by plugging in eq.~(\ref{Kappa}). Let's have a closer look at the QFT in the limit of $(\gamma_{j},\gamma_{k})\to \pm(1,1)$. The boundness $K_{jk}$, and the multiplicative factors $(1\pm \gamma_{j})$ in~(\ref{eq:QFT}) causes each term with $|\gamma_{j}|\to 1$ to vanish. This means that the QFT has a well defined value in the above limit, and we can safely extend by continuity the QTF to the sub-manifolds $(\gamma_{j},\gamma_{k})=\pm(1,1)$.  

The explicit expression of $I_{\mu\nu}$ produces the following results for the Bures metrics
\begin{equation}
g_{\mu\nu}=\frac{1}{4}\Re({I_{\mu\nu})}=\frac{1}{8}\Tr{(K_{\mu}K_{\nu}-\Gamma K_{\mu}\Gamma K_{\nu})}
=-\frac{1}{8}\Tr{(\partial_{\mu}\Gamma K_{\nu})}
=\frac{1}{8}\sum_{jk}\frac{(\partial_{\mu}\Gamma)_{jk}(\partial_{\nu}\Gamma)_{kj}}{1-\gamma_{j}\gamma_{k}},
\end{equation}
which was already derived by Banchi et al.~\cite{Banchi2014}. For the MUC the explicit expression is
\begin{equation}\label{MUCCov}
\mathcal{U}_{\mu\nu}=-\frac{1}{2}\Im({I_{\mu\nu})}=\frac{i}{4}\Tr{(\Gamma[K_{\mu},K_{\nu}])}
=\frac{i}{4}\sum_{jk}\frac{\gamma_{k}-\gamma_{j}}{(1-\gamma_{j}\gamma_{k})^{2}}(\partial_{\mu}\Gamma)_{jk}(\partial_{\nu}\Gamma)_{kj},
\end{equation} 
which, in a parameter-independent way, reads
\begin{equation}
\mathcal{U}=\frac{i}{4}\Tr{(\Gamma K\wedge K)}.
\end{equation}

\section{Sufficient condition for criticality in translationally invariant dissipative models}\label{sec:TIM}
In this section we will show that a singular dependence of $\mathcal{U}$ on the parameters $\lambda\in\mathcal{M}$ necessarily implies a criticality, strictly in the sense of a diverging correlation length.\\\indent  
In a translationally invariant chain, the fermions can be labelled as $\bm{\omega}_{\vv{r}}=(\omega_{r,1},\omega_{r,2})^{T}$, where  $\omega_{r,\beta}$ with $\beta=1,2$ are the two types of Majorana fermions on each site $\vv{r}\in\mathbb{Z}$. The Hamiltonian can be written as $\mathcal{H}=\sum_{\vv{r},\vv{s}}\bm{\omega}^{T}_\vv{r} h(\vv{r}-\vv{s})\bm{\omega}_{\vv{s}}$ and similarly the Lindbladians $\Lambda_{\alpha}(\vv{r})=\sum_{\vv{s}} \bm{l}_{\alpha}^{T}(\vv{s}-\vv{r})\bm{\omega}_{\vv{s}}$, where $h(\vv{r})=h(\vv{r-})^{\dagger}$ are $2\times2$ complex matrices and and $\bm{l}_{\alpha}(\vv{r})\in\mathbb{C}^{2}$. Accordingly, the bath matrix can be expressed as $[M]_{(\vv{r},\beta)(\vv{s},\beta')}=[m(\vv{r}-\vv{s})]_{\beta \beta'}$, ($\beta,\beta'=1,2$), where $m(\vv{r})=m^{\dagger}(-\vv{r})$ are the $2\times2$ matrices $m(\vv{r}):=\sum_{\alpha,\vv{s}}\bm{l}_{\alpha}(\vv{s}-\vv{r})\otimes\bm{l}_{\alpha}^{\dagger}(\vv{s})$.\\\indent
In the limit of infinite large system, both Hamiltonian and bath matrix are circulant, so it is the correlation matrix of the unique steady state solution: $[\Gamma]_{(\vv{r},\beta)(\vv{s},\beta')}=[\gamma(\vv{r}-\vv{s})]_{\beta \beta'}:=1/2\Tr{\rho[\omega_{\vv{r},\beta},\omega_{\vv{s},\beta'}]}$. The Fourier component of the covariance matrix, called the covariance symbol, reads $\tilde{\gamma}{(\vv{\phi})} := \sum_{\vv{r}} \gamma{(\vv{r})}e^{-i\vv{\phi} \cdot \vv{r}}$, with $\vv{\phi}\in [-\pi,\pi)$. As mentioned in the main text, the continuous Lyapunov equation reduces to a set of $2\times2$ matrix equations, labeled by $\phi$
\begin{equation}\label{SCLEphi}
	\tilde{x}(\vv{\phi}) \tilde{\gamma}(\vv{\phi})+\tilde{\gamma}(\vv{\phi})\tilde{x}^{T} (-\vv{\phi})= \tilde{y}(\vv{\phi}),
\end{equation}
 where $\tilde{x}(\vv{\phi})=2[ 2i \tilde{h}(\vv{\phi})+\tilde{m}(\vv{\phi})+\tilde{m}^{T}(-\vv{\phi})]$ and $\tilde{y}(\vv{\phi})=-4[\tilde{m}(\vv{\phi})-\tilde{m}^{T}(-\vv{\phi})]$ are the symbol functions of $X$ and $Y$, respectively, and $\tilde{h}$, $\tilde{m}$ and $\tilde{\bm{l}}_{\alpha}$ are the Fourier components of $h(r)$, $m(r)$ and $\bm{l}_{\alpha}(r)$, respectively. Notice that $\tilde{m}(\phi)=\tilde{m}(\phi)^{\dag}=\sum_{\alpha} \tilde{\bm{l}}_{\alpha}\otimes \tilde{\bm{l}}_{\alpha}^{\dagger}\ge 0$ is a positive semidefinite matrix.

As mentioned in the main text, we can derive a more convenient expression of the MUC by means of the convolution theorem on the third expression of equation~(\ref{MUCCov}). This leads to the following form for the MUC per site
\begin{equation}\label{iota}
\bar{\mathcal{U}}_{\mu\nu}:=\lim_{n\to\infty} \frac{\mathcal{U}_{\mu\nu}}{n}=\frac{1}{(2\pi)}\int_{-\pi}^{\pi}d\phi  \,\,u_{\mu\nu}(\vv{\phi}),
\end{equation}
with 
\begin{equation}\label{Uphi}
u_{\mu\nu}(\phi):=\frac{i}{4}\Tr\{\tilde{\gamma}{(\vv{\phi})}[\kappa_{\mu}{(\vv{\phi})},\kappa_{\nu}{(\vv{\phi})}]\}=\frac{i}{4}\Tr\{\kappa_{\nu}{(\vv{\phi})}[\tilde{\gamma}{(\vv{\phi})},\kappa_{\mu}{(\vv{\phi})}]\},
\end{equation}
where  $\kappa_{\mu}{(\vv{\phi})}$  is the symbol function of $K_{\mu}$, and it is the solution of the $2\times2$ discrete Lyapunov equation
\begin{equation}\label{DLEphi}
\partial_{\mu}\tilde{\gamma}{(\vv{\phi})}=\tilde{\gamma}{(\vv{\phi})}\kappa_{\mu}{(\vv{\phi})} \tilde{\gamma}{(\vv{\phi})} - \kappa_{\mu}{(\vv{\phi})}.
\end{equation}
In the eigenbasis of $\tilde{\gamma}(\phi)$, with eigenvalues $\tilde{\gamma}_{j}$, the explicit solution of~(\ref{DLEphi}) reads 
\begin{equation}
(\kappa_{\mu}(\phi))_{jk}=\frac{(\partial_{\mu}\tilde{\gamma}(\phi))_{jk}}{1-\tilde{\gamma}_{j}\tilde{\gamma}_{k}}. 
\end{equation}
Notice that the diagonal terms $(\kappa_{\mu}(\phi))_{jj}$ provide vanishing contributions to eq.~(\ref{Uphi}) (they commute with $\tilde{\gamma}{(\vv{\phi})}$). Hence, eq.~(\ref{Uphi}) can be cast in the following basis independent form
\begin{equation}\label{Uphi1}
u_{\mu\nu}(\phi) =\left\{ \begin{array}{ll }
      \frac{i}{4}\frac{\Tr\{\tilde{\gamma}{(\vv{\phi})}[\partial_{\mu}\tilde{\gamma}{(\vv{\phi})},\partial_{\nu}\tilde{\gamma}{(\vv{\phi})}]\}}{(1-\Det{\tilde{\gamma}{(\vv{\phi})}})^{2}}& \Det{\tilde{\gamma}{(\vv{\phi})}}\neq 1\\
       0 & \Det{\tilde{\gamma}{(\vv{\phi})}}=1
\end{array} \right. .
\end{equation}
The condition $u_{\mu\nu}(\phi)=0$ for $\Det{\tilde{\gamma}{(\vv{\phi})}}=1$ is a consequence of the continuity argument of $\mathcal{I}_{\mu\nu}$ in $|\gamma_{j}| = 1$, as discussed explicitly at the end of section~\ref{sec:Gauss}. 

Notice that, under the standard isomorphism, $A=a_{jk}\ket{j}\bra{k}\to \text{vec}{(A)}:=a_{jk}\ket{j}\otimes\ket{k}$, the continuous Lyapunov equation~(\ref{SCLEphi}) can be represented by
\[
\hat{X}(z) \V{\tilde{\gamma}(\vv{z})}=\V{y(\vv{z})},
\]
where $\hat{X}(z):=x(z)\otimes\one+\one\otimes x(z^{-1})$. When $\Det{\hat{X}(z)}\neq0$, the unique solution of the symbol function is found simply as 
\begin{equation}\label{sol}
\tilde{\gamma}{(z)}=\frac{\eta{(z)}}{d(z)},
\end{equation}
where $\V{\eta}:=\text{adj}(\hat{X}) \V{y}$, $\text{adj}(\hat{X})$ is the adjugate matrix of $\hat{X}$ and $d(z):=\Det{\hat{X}(z)}$. By construction, $x(z)$ and $y(z)$, and therefore $\eta(z)$ and $d(z)$, are polynomials in $z$ and $z^{-1}$ with coefficients smoothly dependent on system parameters. Hence, $\tilde{\gamma}{(z)}$'s poles are to be found among the roots $\bar{z}$ of $d(z)=0$. Thus, a \emph{necessary} condition for criticality is that, for $\lambda\to\lambda_{0}$, a given root $\bar{z}$ approaches the unit circle $S_{1}$. This clearly means that for $\lambda=\lambda_{0}$, there must exists $\bar{z}_{0}$ such that $|\bar{z}_{0}|=1$ and $\Det{\hat{X}(\bar{z}_{0})}=0$, which implies a vanishing dissipative gap $\Delta:=2 \min_{|z|=1,j}\Re{x_{j}({z})}$, where $x_{j}({z})$'s are the eigenvalues of $\tilde{x}(z)$~\cite{Prosen2010a}.

On the other hand, it may well be the case that all those roots $\bar{z}$ which approach the unit circle as $\lambda\to\lambda_{0}$ are removable singularities of (\ref{sol}). This would result in a finite correlation length, even when $\Delta\to0$. That this may be the case can be readily checked with the example in section~\ref{sec:Example}.

We will next show that a singular behaviour of $\mathcal{U}$ with respect to the parameters is a sufficient condition for criticality. 
First of all, notice, from the equation~(\ref{Uphi1}), that $u(\phi)$ may depend on the dynamics only through $\tilde{\gamma}$, hence any closure of the gap which does not affect the analytical properties of $\tilde{\gamma}$ cannot result in a singular behaviour of $\mathcal{U}$(see also Lemma 2 in the following). We will just need to show that a necessary condition for a singular behaviour of $u(\phi)$ is $\Delta=0$.

Indeed, let's now show that the poles of $u_{\mu\nu}(z)$ on the unit circle $S_{1}$ are to be found only among the roots of $d(z)$. Assuming $d(z)\neq0$, and plugging the unique solution~(\ref{sol}) into equation~$(\ref{Uphi1})$ leads to
\[
u_{\mu\nu}(z) = \frac{N(z)}{D(z)}=\frac{i}{4} \frac{d(z)\Tr\{\eta{(z)}[\partial_{\mu}\eta{(z)},\partial_{\nu}\eta{(z)}]\}}{(d(z)^{2}-\text{Det}{\eta{(z)}})^{2}},
\]
where the numerator $N(z)$ and denominator $D(z)$ are polynomials in $z$ and $z^{-1}$ with smooth dependence on $\lambda$'s. 

We will demonstrate the following: \emph{(i)} that all roots of $d(z)$ lying on the unit circle $S_{1}$ are also roots of $D{(z)}$, and \emph{(ii)} that any other roots of $D(z)$ on $S_{1}$ are not poles of $u_{\mu\nu}(z)$.  For the statement \emph{(i)}, it is just enough to prove the following lemma.\\

\emph{Lemma 1. $d(z)=0$ with $|z|=1$ imply $\eta(z)=0$}.\\
For $z=e^{i\phi}$, $\tilde{x}(\phi)^{\dagger}=\tilde{x}(-\phi)^{T}$, and the eingenvalues of $\hat{X}$ are $x_{j}+x_{k}^{*}$ with $j,k=1,2$, where $x_{j}$  are the eigenvalues of $\tilde{x}$. Since $\Re x_{j}\ge 0$, $\Det{\hat{X}}=0$ implies that $\Delta=2\Re x_{1}=0$, where $x_{1}$ is the eigenvalue of  $\tilde{x}$ with the smallest real part. In~\cite{Prosen2010} it is shown that when $2\Re x_{1}=0$, $\tilde{x}$ is diagonalisable. If $\ket{j}$ is the eigenstate of $\tilde{x}$ with eigenvalue $x_{j}$, then 
$x_{1}+x_{1}^{*}=\bra{1}\tilde{x}(\phi)+\tilde{x}(-\phi)^{T}\ket{1}=4\bra{1}\tilde{m}(\phi)+\tilde{m}(-\phi)^{T}\ket{1}=0$, where in the second equality we used the definition of $\tilde{x}$ and the antisymmetry $\tilde{h}(\phi)=-\tilde{h}(-\phi)^{T}$. 
 From the positive semi-definiteness of the $\tilde{m}$ matrices, it follows that $\bra{1}\tilde{y}(\phi)\ket{1}=-4\bra{1}\tilde{m}(\phi)-\tilde{m}(-\phi)^{T}\ket{1}=0$.

In the eigenbasis $\ket{j}\otimes\ket{k}$, the adjugate matrix is diagonal, and due to $x_{1}+x_{1}^{*}=0$, all elements vanish, except possibly $\bra{1,1}\text{adj}(\hat{X})\ket{1,1}$. On the other hand, the element $\V{\tilde{y}}_{11}=\bra{1}\tilde{y}\ket{1}=0$, implying $\V{\eta}=0$.  \QED

To prove statement \emph{(ii)}, let $\bar{z}_{0}$ be a root of $D(z)$ on $S_{1}$, with $d(\bar{z}_{0})\neq 0$. If $d(z)\neq0$, $\tilde{\gamma}(z)$ in~(\ref{sol}) is the unique solution of the Lyapunov equation~(\ref{SCLEphi}) and it is analytic in $z_{0}$ and smoothly dependent on $\lambda$'s. Since $D(z)=d(z)^{4}[1-\Det\tilde{\gamma}(z)]^{2}$, we obviously have $\Det{\tilde{\gamma}(\bar{z}_{0})}=1$.  To prove statement \emph{(ii)}, we just need the following lemma.

\emph{Lemma 2. If $\tilde{\gamma}(z)$ is analytic in $\bar{z}_{0}\in S_{1}$ and a smooth function of the parameters $\lambda\in\mathcal{M}$, then $u_{\mu\nu}(z)$ is analytic in $z_{0}$.}\\
Just observe that if $\gamma(z)$ is an analytic, smoothly dependent on the parameters, $u_{\mu\nu}(z)$ may be singular in $\bar{z}_{0}$ only if $\Det{\tilde{\gamma}(\bar{z}_{0})}=1$. Assume then $\Det{\tilde{\gamma}(\bar{z}_{0})}=1$, then either $\gamma(\bar{z}_{0})=\pm\one$. Without loss of generality, we can write $\tilde{\gamma}(z)=\one+T \,(z-\bar{z}_{0})^{2n}+ \mathcal{O}(z-\bar{z}_{0})^{2n}$, $n\in\mathbb{N}$, where $T=T^{\dagger}$ is the first non-vanishing term of the Taylor expansion of $\tilde{\gamma}(z)-\one$. The fact that this term must be of even order ($2n$) is due to the positive semi-definiteness of the $\one-\tilde{\gamma}(z) $ for $z\in S_{1}$. By expressing the $2\times2$ matrix $T$ in terms of Pauli matrices, $T=t_{0}\one+ \bm{t}\cdot\bm{\sigma}$, where $\bm{\sigma}:=(\sigma_{x},\sigma_{y},\sigma_{z})^{T}$, $t_{0}\in \mathbb{R}$ and $\bm{t}\in \mathbb{R}^{3}$, the positive semi-definiteness condition above reads: $t_{0}<0$ and $||\bm{t}||\le |t_{0}|$. Plugging the Taylor expansion in~(\ref{Uphi1}) and retaining only the first non-vanishing terms, yields
\[
u_{\mu\nu}(z)=-\frac{1}{4}\frac{\bm{t}\cdot(\partial_{\mu}\bm{t}\wedge\partial_{\nu}\bm{t})}{t_{0}^{2}}(z-\bar{z}_{0})^{2n} + o(z-\bar{z}_{0})^{2n}.\QED
\]

We have thus proven that a non-analycity of $u_{\mu\nu}(z)$ in $\bar{z}_{0}\in S_{1}$ is necessarily due to a pole $\bar{z}$ of $\tilde{\gamma}(z)$ approaching $\bar{z}_{0}$, as $\lambda\to\lambda_{0}$, resulting in a diverging correlation length. Therefore, a singular behaviour of $\bar{\mathcal{U}}$ in the manifold $\mathcal{M}$ is a sufficient criterion for criticality.

\section{A model with closing dissipative gap without criticality}\label{sec:Example}
\begin{figure*}[t]
\mbox{
\subfigure{\includegraphics[scale=0.43]{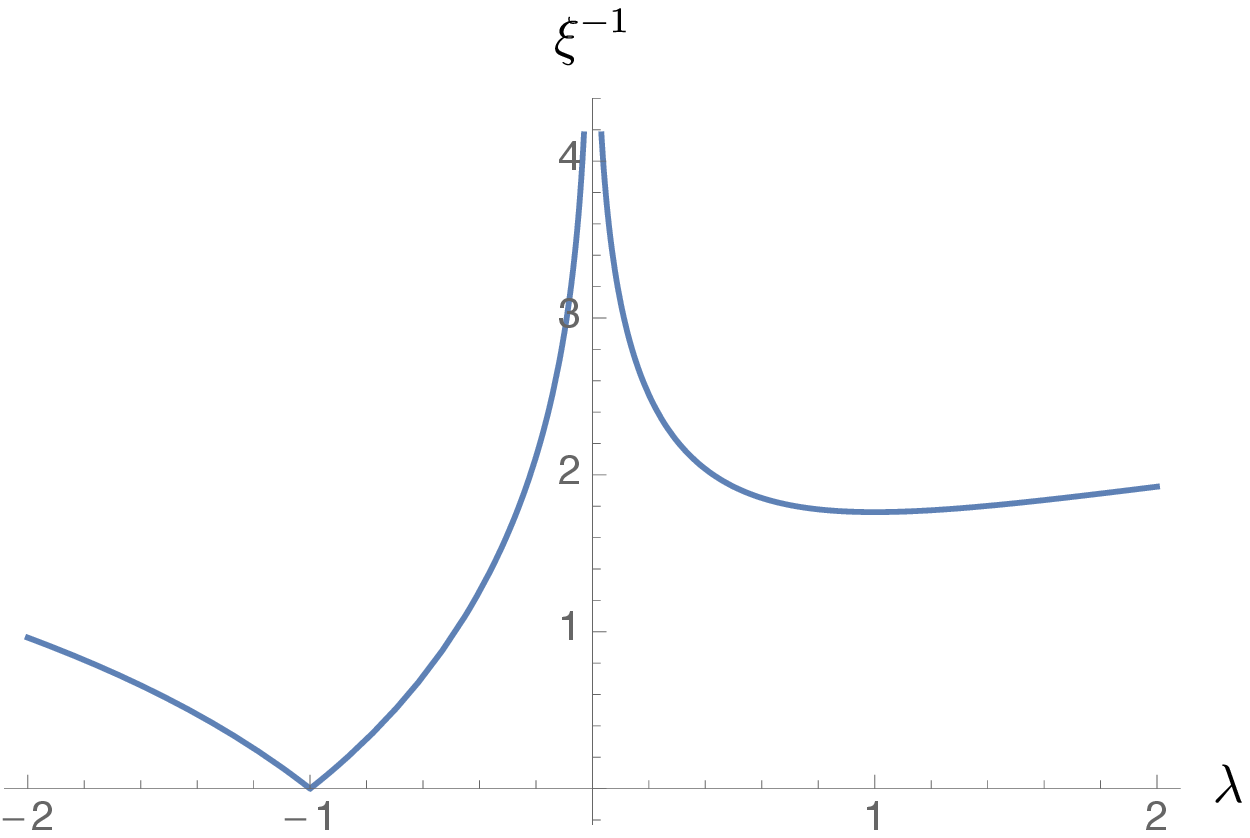}}\qquad
\subfigure{\includegraphics[scale=0.43]{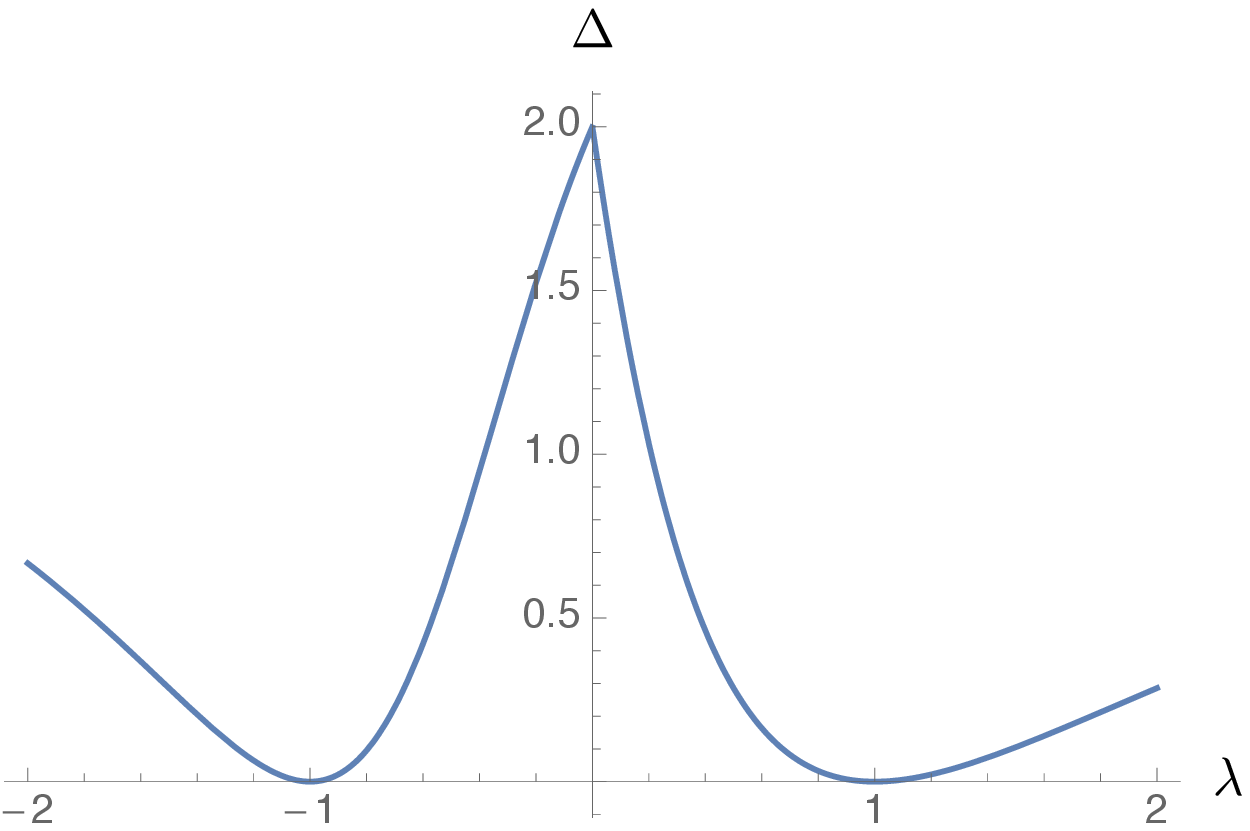}}\qquad
\subfigure{\includegraphics[scale=0.43]{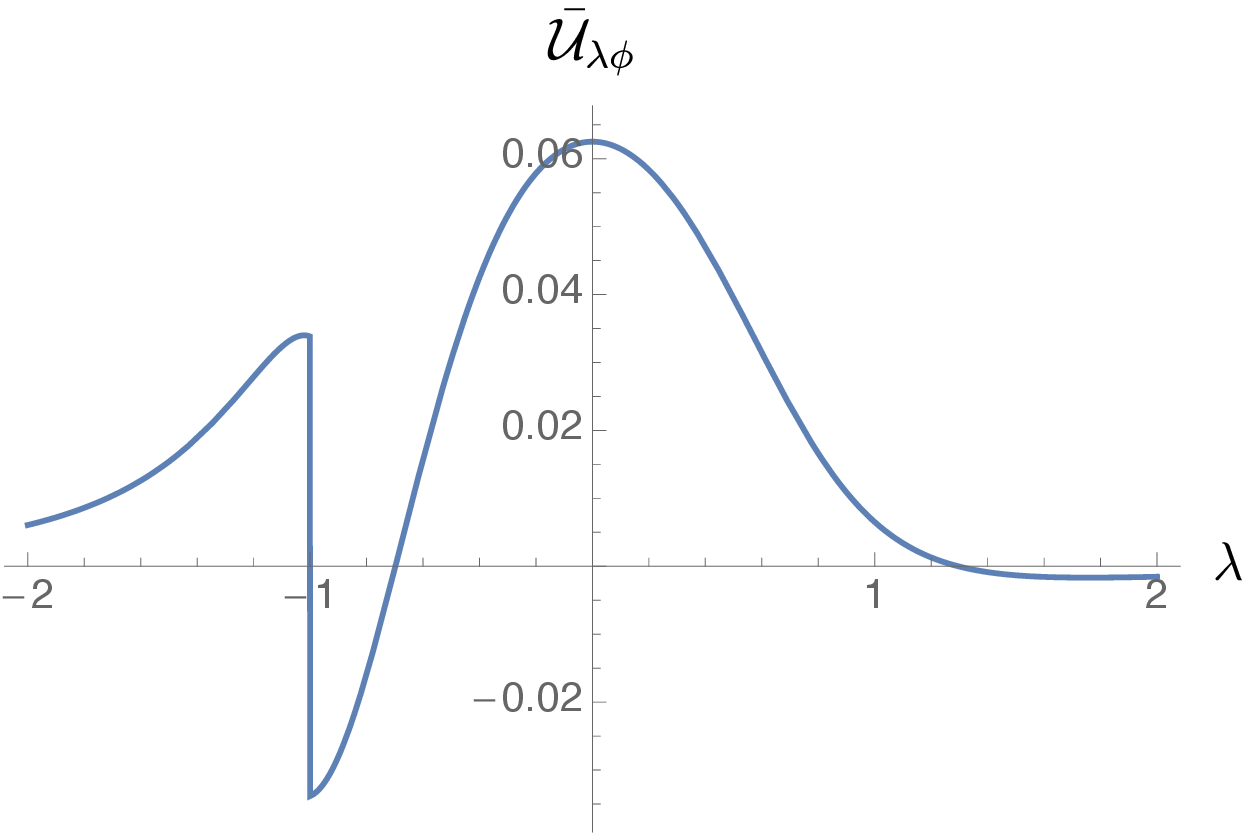}}
}
\vspace{-0.2 cm}
\caption{Model of a 1D fermionic chain on a ring showing a closing dissipative gap that does not imply a diverging correlation length. This is the model discussed in section~\ref{sec:Example} which is a simple extension of a model introduced in~\cite{Bardyn2013}. The inverse correlation length, the dissipative gap and the MUC are shown, respectevely, from the left to the right panel. The model is critical only for $\lambda=-1$, while the gap closes for both $\lambda=\pm 1$. As expected, the discontinuity of MUC captures the criticality, and it is otherwise insensitive to a vanishing gap.}\label{TBD}
\end{figure*}

In this section we will show an example of a 1D fermionic dissipative system in which the closure of the dissipative gap does not necessarily lead to a diverging correlation length. Consider a chain of fermions on a ring geometry, with \emph{no Hamiltonian} and a reservoir defined by the following set of Lindblad operators
\[
\Lambda(r)=[(1+\lambda)\bm{l}_{0}^{T}\bm{\omega}_{r}+\bm{l}_{1}^{T}\bm{\omega}_{r+1}+\lambda\bm{l}_{2}^{T}\bm{\omega}_{r+2}]/n(\lambda),
\]
where  $r=1,\dots,n$, $\bm{l}_{0}=(\cos{\theta},-\sin{\theta})^{T}$, $\bm{l}_{1}=\bm{l}_{2}=i(\sin{\theta},\cos{\theta})^{T}$, and $n(\lambda)=4(\lambda^{2}+\lambda+1)$, with $\lambda\in\mathbb{R}$, $\theta=[0,2\pi)$. %, $\bm{\omega}_{n+1}=\bm{\omega}_{0}$} 
This is a simple extension of a model introduced in~\cite{Bardyn2013}, which, under open boundary conditions, shows a dissipative topological phase transition for $\lambda=\pm1$. In the thermodynamical limit $n\to\infty$, the eigenvalues of $\tilde{x}(\phi)$ are $x_{1}=4(1+\lambda)^{2}/n(\lambda)^{2}$, and $x_{2}=4(1+2\lambda\cos{\phi}+\lambda^{2})/n(\lambda)^{2}$, showing a closure of the dissipative gap at $\lambda=\pm1$. For $|\lambda|\neq 1$ the unique NESS is found by solving the continuous Lyapunov equation~(\ref{SCLEphi}). The symbol function, in a Pauli matrix representation, results $\tilde{\gamma}(\phi)=\bm{\gamma}\cdot\bm{\sigma}$, with
\[
	\bm{\gamma}=g(\phi)\left[\begin{array}{c}(\sin\phi+\lambda\sin{2\phi})\cos 2\theta \\(\cos\phi+\lambda\cos{2\phi}) \\-(\sin\phi+\lambda\sin{2\phi})\sin 2\theta\end{array}\right],
\]
where $g(\phi)=(1+\lambda)/(1+\lambda+\lambda \cos{\phi} + \lambda^{2})$, with eigenvalues $\pm g(\phi)\sqrt{1+\lambda^{2}+2\lambda \cos\phi}$. This shows that $\tilde{\gamma}$ is critical in the sense of diverging correlation, only for $\lambda=-1$ and not for $\lambda=1$, even if the dissipative gap closes in both cases. Figure~\ref{TBD} shows the dependence of the inverse correlation length of the bulk, the dissipative gap and the mean Uhlmann curvature $\bar{\mathcal{U}}_{\lambda \phi}$ on the parameter $\lambda$. Notice a discontinuity of the Uhlmann phase corresponding to the critical point $\lambda_{0}=-1$, while it does not show any singularity for $\lambda=1$ where the gap closes. 
% \begin{figure*}
%    \centering
%    \begin{subfigure}[t]{0.3\textwidth}
%        \centering
%        \includegraphics[width=\linewidth]{TBDCorr.eps}\label{TBD1} 
%        %\caption{ Inverse of correlation length} 
%    \end{subfigure}
%    \hfill
%    \begin{subfigure}[t]{0.3\textwidth}
%        \centering
%        \includegraphics[width=\linewidth]{TBDGap.eps}\label{TBD2}
%        %\caption{Dissipative Gap} 
%    \end{subfigure}
%    \hfill
%    \begin{subfigure}[t]{0.3\textwidth}
%    \centering
%        \includegraphics[width=\linewidth]{TBDuPh.eps}\label{TBD3}
%        %\caption{ MUC.} 
%    \end{subfigure}
%    \caption{Model of a 1D fermionic chain on a ring showing a closing dissipative gap that does not imply a diverging correlation length. The model considered is a simple extension of a model introduced in~\cite{Bardyn2013} discussed in section~\ref{sec:Example}. The inverse correlation length, the dissipative gap and the MUC are shown, respectevely, from the left to the right panel. The model is critical only for $\lambda=-1$, while the gap closes for both $\lambda=\pm 1$. As expected, the discontinuity of MUC captures the criticality, while it is insensitive to the vanishing gap, per s\'e.}\label{TBD}
%\end{figure*}

\section{The Mean Uhlmann Curvature and the Fisher information matrix}
As mentioned in the main text, an important interpretation of $\mathcal{U}$ comes in the context of quantum metrology. The inverse $J^{-1}$ of the Fisher Information matrix (FIM), $J_{\mu\nu}=\frac{1}{2}\Tr\rho\{L_{\mu},L_{\nu}\}$, sets the Cram\'er-Rao bound (CRB)~\cite{Helstrom1976,*Holevo2011,*Paris2009}, i.e. a bound on the estimation precision of the parameters $\lambda\in\mathcal{M}$ labelling a quantum state, i.e.
\begin{equation}\label{eq:SCRB}
\Cov(\hat{\lambda})\ge J^{-1},
\end{equation}
where $\Cov(\hat{\lambda})_{\mu\nu}=\langle (\hat{\lambda}_{\mu}-\lambda_{\mu})(\hat{\lambda}_{\nu}-\lambda_{\nu})\rangle$ is the covariance matrix of a set of locally unbiased estimators $\hat{\lambda}$ of the $\lambda$'s. The expression~(\ref{eq:SCRB}) should be understood as a matrix inequality. In general, one writes 
\[
\tr(G \Cov(\hat{\lambda}))\ge\tr(G J^{-1}),
\]
where $G$ is a given positive definite cost matrix, which allows the uncertainty cost of different parameters to be weighed unevenly. In the case of the estimation of a single parameter $\lambda_{\mu}$, the above inequality can always be saturated, with the optimal measurement protocol being the projective measurement in the eigenbasis of the symmetric logarithmic derivative $L_{\mu}$. However, in the multi-parameter scenario the CRB cannot always be saturated. Intuitively, this is due to the incompatibility of the optimal measurements for different parameters. A sufficient condition for the saturation is indeed $[L_{\mu},L_{\nu}]=0$, which is however not a necessary condition. Within the comprehensive framework of quantum local asymptotic normality (QLAN)~\cite{Hayashi2008,*Kahn2009,*Gill2013,*Yamagata2013}, a necessary and sufficient condition for the saturation of the multi-parameter CRB is given by $\mathcal{U}_{\mu\nu}=0$ for all $\mu$ and $\nu$~\cite{Ragy2016}. \\\indent 
Here, we show explicitly  that $\mathcal{U}_{\mu\nu}$ provides a figure of merit for the discrepancy between an attainable multi-parameter bound and the single parameter CRB quantified by $J^{-1}$.
We will confine ourself to the broad framework of QLAN, in which the \emph{attainable} multi-parameter bound is given by the so called Holevo Cramer-Rao bound (HCRB)~\cite{Helstrom1976,*Holevo2011,*Paris2009}. For a $N$-parameter model, the HCRB can be expressed as~\cite{Hayashi2008} 
\begin{equation}
\tr(G \Cov(\hat{\lambda}))\ge C_{H}(G),
\end{equation}
where 
\begin{equation}
C_{H}(G):=\min_{\{X_{\mu}\}}\{\tr (G \Re Z) +||(G \Im Z)||_{1}\}.
\end{equation}
The $N\times N$ Hermitian matrix is defined as $Z_{\mu\nu}:=\Tr (\rho X_{\mu}X_{\nu})$, where $\{X_{\mu}\}$ is an array of $N$ Hermitian operators on $\HH$ satisfying the unbiasedness conditions $\Tr(\rho X_{\mu})=0$ $\forall \mu$ and $\Tr (X_{\mu} \partial_{\nu}\rho)=\frac{1}{2}\Tr \rho \{X_{\mu}, L_{\nu}\}=\delta_{\mu\nu}$ $\forall \mu,\nu$, and $||B||_{1}$ denotes the sum of all singular values of $B$. %, which in the case of an antisymmetric operator $X$ like $\Im Z$, is equivalent to $||GX||_{1}=\tr\sqrt{G^{1/2} X^{\dagger} G X G^{1/2}}.$ 
If one chooses for $\{X_{\mu}\}$ the array of operators $\tilde{X}_{\mu}:=\sum_{\nu} [J^{-1}]_{\mu\nu} L_{\nu}$, it yields
\begin{equation}
	Z=\tilde{Z}:=J^{-1} I  J^{-1} =  J^{-1} - i 2 J^{-1} \mathcal{U} J^{-1},
\end{equation}  
where $I_{\mu\nu}:=\Tr{\rho L_{\mu} L_{\nu}}$ is the quantum Fisher tensor, and $\mathcal{U}$, with a little abuse of formalism, is the matrix of elements $\mathcal{U_{\mu\nu}}=\frac{i}{4}\Tr\rho[L_{\mu},L_{\nu}]$. If one indicates by $\mathcal{D}(G):=C_{H}(G) - \tr{G J^{-1}}$ the discrepancy $\mathcal{D}(G)$ between the attainable multi-parameter HCRB and the CRB is bounded as follows
\begin{equation}
0 \le \mathcal{D}(G)\le 2 ||G\, J^{-1} \mathcal{U} J^{-1}||_{1},
\end{equation}  
where the first inequality is saturated iff $\mathcal{U}=0$~\cite{Ragy2016}.\\ 
For the special case of a two-parameter model, in the eigenbasis of $J$, with eigenvalues $j_{1}$ and $j_{2}$, it holds
\begin{equation}
J^{-1}\mathcal{U}J^{-1}=\left(\begin{array}{cc}j_{1}^{-1}&0 \\0 & j_{2}^{-1}\end{array}\right)\left(\begin{array}{cc} 0 &\mathcal{U}_{12} \\-\mathcal{U}_{1 2} & 0\end{array}\right) \left(\begin{array}{cc}j_{1}^{-1}&0 \\0 & j_{2}^{-1}\end{array}\right) = \left(\begin{array}{cc} 0 &\frac{\mathcal{U}_{1 2}}{\Det J} \\-\frac{\mathcal{U}_{1 2}}{\Det J} & 0\end{array}\right).
\end{equation}  
It follows that 
\begin{equation}\label{AbsIneq}
2 ||G\, J^{-1} \mathcal{U} J^{-1}||_{1}=2\sqrt{\Det G}\,\,\frac{\sqrt{\Det \,2 \mathcal{U}}}{\Det{J}}.
\end{equation}
Hence, in this case $\sqrt{\Det \,2 \mathcal{U}}/{\Det{J}}$ provides a figure of merit which measures the \emph{amount of incompatibility} between two independent parameters in a quantum two-parameter model. \\\indent  
For self-adjoint operators $B_{1},\dots,B_{N}$, the Schrodinger-Robertson's uncertainty principle is the inequality~\cite{Robertson1929}
\begin{equation}
	\Det\left[\frac{1}{2} \Tr \rho \{B_{\mu},B_{\nu}\}\right]_{\mu,\nu=1}^{N}\ge\Det\left[-\frac{i}{2} \Tr\rho[B_{\mu},B_{\nu}]\right]_{\mu,\nu=1}^{N},
\end{equation}
which applied to the SLD $L_{\mu}$'s, yields
\begin{equation}\label{DetIneq}
	\Det J \ge \Det 2\, \mathcal{U}.
\end{equation}
For $N=2$,  when the inequality~(\ref{DetIneq}) \label{AbsIneq} is saturated, it implies that
\begin{equation}\label{AbsIneq}
\mathcal{D}(G) \simeq 2\sqrt{\Det G J^{-1}},
\end{equation}
which means that the discrepancy $\mathcal{D}(G)$ reaches the same order of magnitude of  $\tr(G J^{-1})$, i.e. the CRB itself. This limit marks the \emph{condition of maximal incompatibility} for the two-parameter estimation problem, arising from the quantum nature of the underlying system.\\ 
Another interesting inequality relates the eigenvalues of $J$ and $\mathcal{U}$. The QFT $I_{\mu\nu}=\Tr\rho L_{\mu}L_{\nu} = J_{\mu\nu}-i2\mathcal{U}_{\mu\nu}$ is a positive (semi)-definite Hermitian matrix. Hence, by definition $J\ge i 2\, \mathcal{U}$, in a matrix sense. It follows that~\cite{Horn2013}
\begin{equation}
	j_{i}\ge 2 i\, u_{i},
\end{equation}
where $j_{i}$ and $u_{i}$ are the i-th eigenvalues of $J$ and $\mathcal{U}$, respectively, ordered according to $j_{1}\le j_{2}\le\dots\le j_{N}$ and $u_{1}\le u_{2}\le\dots\le u_{N}$.  In particular, for i=1, one gets
\begin{equation}
	|| J ||_{\infty}\ge 2 || i\, \mathcal{U}||_{\infty}.
\end{equation}

%In general, the SLD, in the eigenbasis of a density matrix $\rho=\sum_{j} p_{j}\ket{j}\bra{j}$, can be expressed as:
%\begin{equation}
%	L_{\mu} = L_{\mu}^{C}+L_{\mu}^{Q},   
%\end{equation}
%where $L_{\mu}^{C}:=2 \sum_{j} \partial_{\mu}(\ln p_{j})\ket{j}\bra{j}$ is the diagonal part of $L_{\mu}$, and accounts for the variations of the eigenvalues of the density matrix, whereas $L_{\mu}^{Q}:=2 \sum_{jk} \frac{p_{k}-p_{j}}{p_{k}+p_{j}}\ket{j}\bra{j}\partial_{\mu}\ket{k}\bra{k}$ is off-diagonal and depends purely on the change of eigenstates of $\rho$. 
%The Fisher information matrix can be expressed into $J=J^{C}+J^{Q}$, the so called classical and quantum Fisher information matrices, respectively, where
%\begin{equation}
%	J^{Q}_{\mu\nu} := \frac{1}{2} \Tr \rho \{L^{Q}_{\mu},L^{Q}_{\nu}\},   
%\end{equation}
%and similarly for $J^{C}$. The Schrodinger-Robertson's uncertainty principle, can readily be applied to the $L^{Q}_{\mu}$'s, yielding:
%\begin{equation}
%	\Det J^{Q} \ge \Det 4\, \mathcal{U}.
%\end{equation}
%And similarly,
%\begin{equation}
%	|| J_{Q} ||_{\infty}\ge 4 || i\, \mathcal{U}||_{\infty}.
%\end{equation}

\bibliography{\bibliopath library}
\end{document}